\documentclass[ba]{imsart}
\pubyear{0000}
\volume{00}
\issue{0}
\doi{0000}
\arxiv{2205.05955}
\firstpage{1}
\lastpage{1}
\usepackage{marginnote}
\usepackage[normalem]{ulem}
\usepackage{enumitem}
\usepackage{amsthm}
\usepackage{multirow}
\usepackage{amsmath}
\usepackage{natbib}
\usepackage[colorlinks,citecolor=blue,urlcolor=blue,filecolor=blue,backref=page]{hyperref}
\usepackage{graphicx}
\usepackage{cleveref}
\usepackage{xcolor}
\usepackage{bm}
\DeclareMathOperator*{\argmin}{arg\,min} 

\startlocaldefs
\numberwithin{equation}{section}
\theoremstyle{plain}

\endlocaldefs

\def\bepsilon{\bm \epsilon}
\def\btheta{\bm \theta}
\def\bkappa{\bm \kappa}
\def\bmu{\bm \mu}
\def\bSigma{\bm \Sigma}
\def\bphi{\bm \phi}

\def\k0{k_*}

\def\phi{\varphi}

\def\DKL{D_\text{KL}}

\def\TNFA{{TNF$\alpha$} }
\def\NFKB{NF-$\kappa$B }
\def\IKBA{I$\kappa$B$\alpha$ }

\def\vec{\text{vec}}
\def\Sy{\underline{\mathbf{s}}}

\def\Om1{\Omega^{-1}}

\def\phi{\varphi}

\newcommand{\gma}[1]{{\color{black} #1}} 

\def\cI{\mathcal{I}}

\def\bS{S}
\def\dth{\delta{\bm \theta}}
\def\th{{\bm \theta}}
\def\cN{\mathcal{N}}

\def\ba{{\bm a}}

\def\be{{\bm e}}

\def\bom{{\bm m}}
\def\bxi{\bm \xi}
\def\bw{\bm w}
\def\br{{\bm r}}

\def\bu{{\bm u}}
\def\bv{{\bm v}}
\def\bx{{\bm x}}
\def\by{{\bm y}}

\def\bA{{\bm A}}
\def\bB{{\bm B}}
\def\bC{{\bm C}}
\def\bE{{\bm E}}
\def\bF{{\bm F}}
\def\bI{{\bm I}}
\def\bJ{\bm J}

\def\bR{{\bm R}}
\def\bS{{\bm S}}

\def\bU{{\bm U}}
\def\bV{{\bm V}}
\def\bY{{\bm Y}}

\def\bphi{{\bm \phi}}

\def\TNFA{TNF-$\alpha$ }
\def\NFKB{NF-$\kappa$B }
\def\IKBA{I$\kappa$B$\alpha$ }

\def\k0{k_*}

\def\phi{\varphi}

\def\DKL{D_\text{KL}}

\def\TNFA{{TNF$\alpha$} }
\def\NFKB{NF-$\kappa$B }
\def\IKBA{I$\kappa$B$\alpha$ }

\def\vec1{\text{vec}}
\def\Sy{{\bm U}} 

\def\NFKB{NF-$\kappa$B }
\def\S0{\bS_0}
\def\cI{\mathcal{I}}
\def\boeta{{\bm \eta}}

\usepackage{xr-hyper}
\makeatletter

\newcommand*{\addFileDependency}[1]{
\typeout{(#1)}
\@addtofilelist{#1}
\IfFileExists{#1}{}{\typeout{No file #1.}}
}\makeatother

\newcommand*{\myexternaldocument}[1]{%
\externaldocument{#1}%
\addFileDependency{#1.tex}%
\addFileDependency{#1.aux}%
}

\myexternaldocument{supplementary}

\usepackage{xr}
\begin{document}

\begin{frontmatter}
\title{Bayesian inference for stochastic oscillatory systems using the phase-corrected Linear Noise Approximation\thanksref{T1}}
\runtitle{Bayesian inference for oscillatory systems}

\begin{aug}
\author{\fnms{Ben} \snm{Swallow}\thanksref{addr1,addr2,t1,t2,m1}\ead[label=e1]{bts3@st-andrews.ac.uk}},
\author{\fnms{David A.} \snm{Rand}\thanksref{addr2,t1,m2}%
\ead[label=e2]{d.a.rand@warwick.ac.uk}}
\and
\author{\fnms{Giorgos} \snm{Minas}\thanksref{addr3,t3,m1,m2}\ead[label=e3]{gm256@st-andrews.ac.uk}}

\runauthor{B. Swallow et al.}

\address[addr1]{School of Mathematics and Statistics, University of St Andrews, St Andrews, UK,   
    \printead{e1}, 
}

\address[addr2]{Mathematics Institute, University of Warwick, Coventry, UK
    \printead{e2}
}

\address[addr3]{School of Mathematics and Statistics, University of St Andrews, St Andrews, UK,
\printead{e3}}


\end{aug}

\begin{abstract}
Likelihood-based inference in stochastic non-linear dynamical systems, such as those found in chemical reaction networks and biological clock systems, is inherently complex and has largely been limited to small and unrealistically simple systems. 
Recent advances in analytically tractable approximations to the underlying conditional probability distributions enable long-term dynamics to be accurately modelled, and make the large number of model evaluations required for exact Bayesian inference much more feasible. 
We propose a new methodology for inference in stochastic non-linear dynamical systems exhibiting oscillatory behaviour \gma{that can be applied even if the system involves a large number of variables and unknown parameters}. 
and show the parameters in these models can be realistically estimated from simulated data.
\gma{We use }preliminary analyses based on the Fisher Information Matrix of the model \gma{to} guide the implementation of Bayesian inference.
We show that this parameter sensitivity analysis can predict which parameters are practically identifiable.
A parallel tempering algorithm 
\gma{is used to provide the flexibility required to explore the posterior distribution of the model parameters, which often exhibit multi-modal posterior distributions.} 

\end{abstract}



\end{frontmatter}

\section{Introduction}
Oscillations are abundant in biology, ecology, epidemiology, and other applied fields \citep{Goldental2017}. 
Examples include genetic oscillations observed in biological systems \citep{Forgerbook,Gonze2021} such as the circadian clock \citep{drosmodel}, embryonic development \citep{Marinopoulou2021}, cell signalling \citep{ashall}, predator-prey oscillations in ecology \citep{gard_kannan_1976,froda07}, and epidemic oscillations \citep{Greer2020,Weitz2020}.
Dynamical systems presenting oscillations naturally carry more information than stable systems through their period, phase, and amplitude components. 
To capture this information, one needs to observe their dynamics over time, rather than through a single static observation. 
time series data are therefore important for prediction and for inference on the parameters of oscillatory models.  
Cutting-edge technology in molecular biology, such as fluorescent imaging and sequencing \citep{Gabriel2021,Lane2017,DeFelice2019}, ecology \citep{mccrea2023}, wearable devices (e.g. GPS, cameras) \citep{spitschan2024}, and widespread surveillance of seasonal epidemics, and elsewhere, provide such time series observations.

Traditionally oscillations are modelled by deterministic dynamical systems, often described by ordinary differential equations. 
However, a key component of oscillatory systems is that they are often stochastic by nature \citep{Boettiger2018,ashall,drosmodel,Allen2017}.
For instance, systems in molecular biology evolve through biochemical reactions affected by the stochastic movement of molecules in biological cells \citep{gillespe,vankampen92,wilkinsonsmsb,sysbio}. 
Similarly predator-prey interactions, and transitions between states (e.g.\ susceptible to infected) in epidemics are highly stochastic \citep{Allen2017}.
This intrinsic stochasticity causes phase diffusion with time series data of oscillatory systems appearing asynchronous even after the first oscillation
\citep{ashall,Gabriel2021}.
Deterministic models fail to capture the intrinsic stochasticity of oscillatory systems providing poor fit to data and poor parameter estimation \citep{ashall,Harper2018a,Tay2010}.

Stochastic models are therefore necessary for parameter estimation using highly-variable time series data. 
These are continuous-time Markov processes, which are typically non-homogeneous as the transition propensities depend on the current state of the system. 
A range of approaches for stochastic modelling are available. 
The model that is considered to be exact (under generic assumptions) in the context of biochemical reactions is the so-called Chemical Master Equation (CME) \citep{Gillespie1992,wilkinsonsmsb,sysbio}.
The well-known Stochastic Simulation Algorithm (SSA), also called the Gillespie algorithm, allows for exact simulation from the CME model and it is widely used for stochastic simulation in many fields \citep{Gillespie1992}. 
However, the likelihood of the CME model is analytically intractable in most situations.  \citep{beaumontabc,abc_book,wilkinsonsmsb}. 
Therefore the use of 
likelihood-free methods, such as Approximate Bayesian Computation (ABC) \citep{beaumontabc,abc_book,wilkinsonsmsb} or
particle Markov Chain Monte Carlo \citep{Golightly2011}, is required to perform parameter inference. 

Less progress has been made in parameter estimation when the dynamics of the system are non-linear (for instance oscillatory), despite there being a wide variety of approximate models enabling faster simulation \citep{Gillespie2003,Gillespie2000a}. 
One such method, the Linear Noise Approximation (LNA) of the CME (\cite{vankampen92,kurtz70}), described by Stochastic Differential Equations (SDEs), provides analytically tractable likelihood computation of time series observations (\cite{Komorowski2009,fearnhead14,minasrand17,Schnoerr,finkenstadt13,girolami11}), but fails to accurately model oscillators (\cite{minasrand17,Ito2010}). 
An extension of the LNA, called phase corrected LNA or pcLNA \citep{minasrand17}, 
can accurately simulate oscillatory dynamics of the CME.
The pcLNA model takes advantage of the linear stability of oscillatory systems in all but the tangential direction of the oscillations while applying frequent phase corrections that capture the observed phase diffusion.
The pcLNA framework provides simulation algorithms that remain accurate for long simulated trajectories and are fast in implementation, thus making likelihood-based parameter estimation methods such as Markov chain Monte Carlo (MCMC) feasible.
The method also allows for computing information theoretic quantities such as the Fisher Information to study the model's parameter sensitivities \citep{Minas2019}, quantities that can be vital for efficient and reliable parameter estimation in such complex systems.

An important characteristic of oscillatory systems is that they typically involve a large number of variables and parameters.
For instance, the model of the circadian clock of \emph{Drosophila.Melanogaster} \citep{Gonze2021} and the NF-$\kappa$B signalling system \citep{ashall} that we consider here involve 10 and 11 variables, and 39 and 30 parameters, respectively. 
These systems describe the biochemical evolution of certain molecular populations over time. 
The transitions include reactions (e.g.\ transcription, translation, degradation, and phosphorylation) and translocations, and the parameters to be estimated include constants describing the speed of reactions, and threshold values for non-linear reactions. 
It is well-known that deterministic models for these systems face substantial parameter identifiability issues (``sloppiness'') in practice \citep{Gutenkunst:2007,Rand2008,Minas2019,browning2020}, with further inferential challenges emerging by stochastic models due to difficulty in estimating joint probability densities of state vectors and parameters. 

This paper 
\gma{develops and applies a method for} Bayesian 
estimation 
\gma{of the parameters of} stochastic, oscillatory dynamical systems 
\gma{using} time series observations.
\gma{For this, we first describe a Kalman Filter that can be used to derive the likelihood of time series observations under the proposed pcLNA model. 
We then perform a parameter sensitivity analysis using the Fisher Information matrix computed under the pcLNA model. 
We finally apply a Parallel Tempering MCMC (PTMCMC) algorithm \citep{ptMCMCclassic} using the pcLNA Kalman-Filter to estimate the posterior distribution of model parameters given time series observations.}

We find that 
\gma{we are able to accurately estimate a number of parameters of large oscillatory systems. 
We also find that}
the results of the parameter sensitivity analysis are reflected in the properties of the parameters' Markov chains, in the sense that parameters that are suggested as non-identifiable in our study are seen to have issues of non-convergence and poor mixing, or at best minimal posterior concentration relative to the prior.
This, to a large extent, is regardless of the \gma{parameters or the} computational method used.
In contrast, the parameters that are identifiable according to our study, appear to converge and mix well in the MCMC methods. 
Overall, the parallel tempering method is found to perform 
well, in terms of \gma{mixing and} convergence, particularly dealing with multi-modal distributions. 
We discuss the selection of various parameters of the PTMCMC algorithm, including proposal distributions, and number of chains.

Previous studies \citep{Komorowski2009,finkenstadt13,girolami11} have examined the use of Bayesian methods to estimate parameters of the LNA model \gma{in simpler settings (e.g.\ linear systems, lower noise levels)}. 
Here we consider the use of those methods for oscillatory dynamics where the LNA model is inaccurate.  
\cite{fearnhead14} uses a different approach in improving the accuracy of the LNA, by applying frequent corrections to the initial conditions of the ODEs used to solve the LNA system equations forward in time. 
As we explain in section \ref{sec:pcLNA}, the pcLNA method takes advantage of the oscillatory dynamics by correcting only the time/phase of a pre-computed solution of the LNA equations, which leads to a significant decrease in the implementation time \gma{(see section \ref{sec:restartLNA})}. The pcLNA approach\gma{, unlike the LNA method in \cite{fearnhead14},} also allows for the computation of the Fisher Information without significant increase in computational time compared to the standard LNA model. Therefore, a parameter sensitivity analysis based on Fisher Information can be performed and in turn guide parameter estimation of oscillatory systems. 

\subsection{Exemplar systems}\label{sec:exsys}
In order to test the methodology, we apply the methods developed in this paper to two biological systems. 
Both of these systems present oscillations generated by negative feedback loops with large numbers of parameters and variables.  

\begin{figure}
    \centering
    \includegraphics{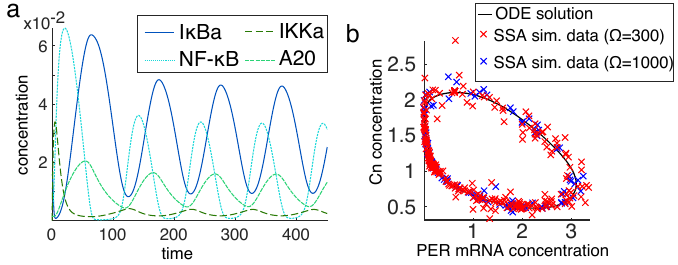}
    \caption{The dynamics of the exemplar systems. a. The (deterministic) solutions of the \NFKB system responding to a continuous \TNFA stimulation plotted against time. Here only 4  variables are presented, specifically the concentrations of the inhibitor protein \IKBA in the cytoplasm, the \NFKB concentration in the nucleus, the concentration of the active kinase of the inhibitor IKKa, and of the A20 protein. Observe that the \IKBA is negatively correlated with \NFKB, same with A20 and IKKa. b. The ODE solutions of the Drosophila circadian clock and simulated data produced using the SSA. Two variables are presented and plotted against each other. These are the mRNA expression levels of PER gene, and the concentration of its inhibitor, $C_n$, which is a a complex formed by a (phosphorylated) form of proteins Per and Tim.}
    \label{fig:detsols}
\end{figure}

Our first exemplar system is that of the NF-$\kappa$B signalling system in mammalian cells. 
The NF-$\kappa$B system regulates cell response to stress and inflammation,
and its dysregulation plays a key role in autoinflammatory diseases, cancer, and other pathological conditions \citep{Zhang2017d}.
We utilise an 11-dimensional system  corresponding to a reduced version of the model of \citet{ashall} as described in \citet{minasrand17}. 
It describes the oscillatory response of the \NFKB system following stimulation by the cytokine tumor necrosis factor alpha (TNF$\alpha$). 
Continuous stimulation of the system causes a transient oscillation that quickly relaxes to a stable limit cycle (see Figure \ref{fig:detsols}a). 
The \NFKB system has been shown to be highly stochastic \citep{Tay}.
The \NFKB model used here
involves reactions modelled by both linear and non-linear rate functions, and has a total of 31 parameters including reaction rate constants, and Michaelis-Menten or Hill equation constants. 

We also consider the circadian clock model for the species \emph{Drosophila.Melanogaster} developed in \citet{drosmodel}. 
The system involves $30$ reactions parameterised by $38$ parameters of similar types as the \NFKB model.
The system is shown to present sustained oscillations generated by negative feedback loops (self-inhibition of gene expression, see Figure \ref{fig:detsols}b).
\citet{Minas2019} analysed this system and showed a substantially higher parameter sensitivity of the stochastic model  described below, compared to the deterministic model of this system. 
Higher relative sensitivity is presented in this system compared to the \NFKB system as described below. 

\section{Reaction network dynamics and the pcLNA model}\label{sec:pcLNA}

We consider a system of multiple populations, $M_1,M_2,\dots,M_n$, that interact with each other according to a fixed number of $R$ interactions, in this context called reactions. 
These populations might be for instance the Susceptible, Infected and Recovered individuals in epidemiology, the predators and prey in ecology, or various molecular populations in biochemistry. 
Reaction $j$ can be described by 
\[ k_{1j} M_1 + \dots + k_{nj} M_n \stackrel{c_j}{\longrightarrow} k_{1j}' M_1 + \dots + k_{nj}' M_n, \qquad k_{lj},k_{lj}'\geq 0, \; l=1,\dots,n \] 
where on the left hand side, $k_{lj}$ gives the number of individuals of population $M_l$ that is needed for the reaction to happen and on the right hand side, $k'_{lj}$ gives the number of $M_l$ individuals that is produced by the reaction, where $l=1,\dots,n$ and $j=1,\dots, R$. 
The populations $M_l$ with $k_{lj} \neq 0$ are called reactants while those with $k'_{lj} \neq 0$ are called products of the reaction.
The vector $\ba_j = (k_{1j}-k_{1j}',\dots,k_{nj}-k_{nj}')^T$\footnote{We write all vectors as column vectors here. 
We denote vectors with bold lower case letters, and matrices with bold upper case letters. The superscript $^T$ denotes the transpose of a vector or matrix.
}
 describes the net change after one occurrence of the $j$-th reaction. We call the matrix $\bA$ with columns $\ba_j$, $j=1,\dots,R$, the stoichiometry matrix of the reaction network. 
The positive constant $c_j$ indicates the rate constant of the reaction. 

We define the state $\bx(t) = (x_1(t),\dots,x_n(t))^T$ where the $l$-th entry gives the number of individuals of the $M_l$ population at time $t\geq 0$. 
Under the assumption that the populations live in a well-mixed environment with constant conditions (see \cite{Gillespie1992}), the occurrence of the next reaction of type $j$ follows an exponential distribution. 
The rate of this exponential distribution depends on the state of the reactants and the constant $c_j$. 
If the current state is $\bx(t)=\bx$, then a general form of the rate of the $j$-th reaction is
$ w_j(\bx) = c_j\prod_l \bx_l^{k_{lj}},$
but it is worth noting that other forms (e.g. $w_j(\bx) = c_j\frac{k}{k + x_l}$, $k>0$) are also commonly used. 
The parameters involved in the reactions (e.g. $c_j$, $k$) are typically unknown and we will attempt to estimate them in the following sections.  
The stochastic process describing the evolution of state $\bx(t)$ in time $t\geq 0$ is a continuous time-inhomogeneous Poisson process, where the inhomogeneity is due to the non-constant rates that depend on the dynamic state of the system.
The evolution of the probability distribution of this Poisson process over time can be described by the so-called master equation, which is the Kolmogorov-Smirnov equation of this process  \citep{wilkinsonsmsb}. 
The master equation can only be solved very rarely, but an exact stochastic simulation called Stochastic Simulation Algorithm (SSA), also known as Gillespie algorithm \citep{Gillespie2000a}, can be used to generate stochastic trajectories that exactly follow the probability distribution of the Poisson process. 
A number of approximate models and simulation algorithms have been developed \citep{Gillespie2003,Gillespie2000a}.

\subsection{The Linear Noise Approximation}\label{sec:LNA} We 
next describe the standard Linear Noise Approximation (LNA) model \citep{vankampen92,kurtz70,kurtz71}. 
The LNA gives the state of the system at time $t>0$ by
\begin{equation}\label{eq:LNAansantz}
	\bx(t) = \Omega\bphi(t) + \sqrt{\Omega}\, \bxi_t.
\end{equation}

The term $\Omega$ is a parameter, 
often called system size, that is typically considered as a known property of the system (e.g.\ cell volume, total population size). 
It is not necessary to identify such a parameter (i.e.\ it can be fixed to 1), but in some cases, in particular for the \NFKB system that we consider, the parameter has been estimated for certain cell types \citep{minasrand17,wilkinsonsmsb}.

Then $\bphi_t = \bphi(t;\btheta)$ is a (deterministic) solution of the Ordinary Differential Equation (ODE)
\begin{equation}\label{eq:ode}
    d\bphi_t = \bF(\bphi_t){dt},\quad  \text{where } \bF(\bphi_t)= \bA \br(\bphi_t)
\end{equation}
 and $\bA$ the stoichiometry matrix  with columns $\ba_j = (k_{1j}-k_{1j}',\dots,k_{nj}-k_{nj}')^T$. The vector $\br(\bphi_t)=(r_1(\bphi_t),\dots,r_R(\bphi_t))^T$ has entries the classical rates of each reaction that satisfy the macroscopic law of reactions (see \cite{wilkinsonsmsb}). 
Each entry $r_j(\cdot)$ is related to the corresponding function $w_j(\cdot)$ typically by $w_j(\bx) = \Omega r_j(\bx/\Omega)$.
The solution $\bphi_t$ is a macroscopic ($\Omega \to \infty$) limit of $\bx(t)/\Omega$.

Finally $\bxi_t=\bxi(t)$ is the solution of the Stochastic Differential Equation (SDE)
\begin{equation}\label{eq:LNAsde}
	d\bxi_t = \bJ_t \bxi_t dt  + \bS_t^{1/2} d\bw_t.
\end{equation}
    The drift matrix of the SDE in \eqref{eq:LNAsde} is the Jacobian matrix of the system in \eqref{eq:ode}, i.e.\ $\bJ_t = \bJ(\bphi_t)= (dF_i(\bphi_t)/d\phi_j)_{i,j}$, while $\bS_t = \bA \bR(\bphi_t)\bA^T$
    with $\bR(\bphi)$ a diagonal matrix with main diagonal $\br(\bphi)$. 
    The stochastic process $\{\bw_t:t\geq 0\}$ is a classical $N$-dimensional Wiener process.
    
The key advantage of the LNA over other approximate models is that the SDE in \eqref{eq:LNAsde} can be solved analytically
with the solution satisfying
\begin{equation}\label{eq:xi}
    \bxi_t = \bC(t_0,t) \bxi_{t_0} + {\bm \eta}_t , \quad {\bm \eta}_t \sim N(0,\bV(t_0,t)).
\end{equation}
Here the matrices $\bC$ and $\bV$ are solutions of the initial value problems, 
\begin{align}
   d\bC/dt&= \bJ_t \bC
       , & \bC(t_0,t_0)=\bI, \label{eq:Code}\\
	\quad 
	d\bV/dt &= \bJ_t \bV + \bV \bJ_t^T + \bS_t, & \bV(t_0,t_0)={\bm 0} \label{eq:Vode}
\end{align}
	 where $\bI$ and ${\bm 0}$ are the $n\times n$ identity and zero matrix, respectively.

Equation \eqref{eq:xi} implies that the transition probabilities $P(\bxi(t_{1}) | \bxi(t_0)=\bxi_0)$ and $P(\bx(t_{1}) | \bx(t_0)=\bx_0)$, 
where $\bxi_0$ and $\bx_0$ are fixed values, are multivariate normal (MVN), for any $t_1>t_0$, and also more generally that, if the initial condition $\bxi(t_0)$ is a random vector with MVN distribution, the transition probabilities of the LNA model
$P(\bx(t_{1}) | \bx(t_0))$
are also MVN. 
We can also show that the joint probability distribution of time series 
$(\bx(t_0),\bx(t_1),\dots,\bx(t_n))$
are multivariate normal with precision matrix that has a block tridiagonal form (see \citep{Minas2019}).
Practically, using the LNA involves solving the ODE system in \eqref{eq:ode} for appropriate initial conditions, and then using this solution $\bphi(t)$ to compute the matrices $\bJ=\bJ(\bphi(t))$ and $\bS=\bS(\bphi(t))$, and solve \eqref{eq:Code}, and \eqref{eq:Vode}. The solutions $\bphi(t)$, $\bC(t,t_0)$, $\bV(t,t_0)$, $t\in[t_0,T]$ can be used for deriving the probability distribution of the state of the system, performing simulations, computing the FIM of the model parameters, statistical inference (see below), and other purposes. 

The accuracy of the standard LNA model
depends on the system dynamics, and the relation between $\Omega$ and $T-t_0$, i.e.\ the length of the time-interval to be described. 
For instance, the LNA model is shown to be accurate for long-times when describing the dynamics of a stochastic dynamical system that has reached the neighborhood of an equilibrium of the system in \eqref{eq:ode}.
An accurate LNA can also be derived for any system when for a given $\Omega$, the length of the time-interval is chosen to be sufficiently short (see \cite{Grima2011}, \cite{Wallace2012}). 
However, it is inaccurate in describing the long-time behaviour of stochastic systems presenting oscillations. 
For instance, we found that the standard LNA model is inaccurate when describing the \NFKB and circadian clock oscillations even for time-intervals less than one oscillation cycle (\cite{minasrand17}).

\subsection{The phase corrected LNA method}\label{sec:pcLNA} 

By studying the dynamics of oscillatory systems that present attractive limit cycless, the phase corrected LNA model (pcLNA) takes advantage of the stability of these systems in all-but-one direction of the state space. 
That is, if we consider a system of the form \eqref{eq:ode} that has a solution that forms an attractive limit cycle, the dynamics of these systems are stable in the $N-1$ transversal directions at any arbitrary point of the limit cycle. 
This stability is reflected into the stochastic dynamics. 
More specifically, \cite{minasrand17} compare the joint probability distributions of SSA trajectories at multiple transversal sections with the corresponding Multivariate Normal distributions derived analytically under the LNA model to show that they are approximately equal.
The variation in the unstable tangental direction can be controlled by frequently correcting the phase of the system between standard LNA transitions. 
The frequency depends on the dynamics of the system and the scale of stochasticity (here expressed by parameter $\Omega$). \cite{minasrand17} observed that about 3-4 corrections per oscillation cycle was sufficient for the \NFKB system and the \emph{Drosophila} circadian clock model ($\Omega \in (300,2000)$). 

\begin{figure}[h]
\centering
\includegraphics{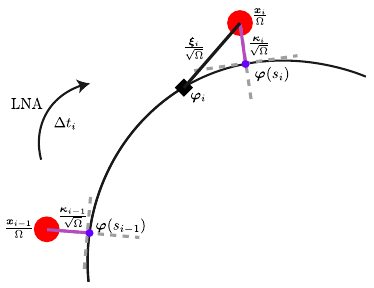}
\caption{\label{fig:pcLNA} The pcLNA transitions. The phase corrected states of the underlying deterministic and stochastic process $\bphi_t$ and $\bxi_t$, respectively, are used to proceed using the standard LNA transition equations.}
\end{figure}

For a system with limit cycle solution, $\bphi(t)$, $t\in[0,\pi]$, with period $\pi$, the pcLNA model at times $t_1,\dots,t_N$ is described by the following equations, which are also visualised in Figure \ref{fig:pcLNA}:
\begin{eqnarray}
\text{(State)} & \bx_{i} =   \bx(t_{i}) =\Omega \bphi_{i}\! +\! \sqrt{\Omega}\, \bxi_{i},& \qquad \label{eq:state}\\
 \text{(Transition)} & \bphi_{i} = \bphi(s_{i-1}\! +\!\Delta t_{i}), & \Delta t_{i} = t_{i}-t_{i-1}\qquad \label{eq:detTr}\\
 &\hspace{-1.5cm} \bxi_{i} = \bC(s_{i-1},s_{i-1}\! +\!\Delta t_{i})\, \bkappa_{i-1}+ \boeta_{i} ,  &\hspace{-0.25cm} \boeta_{i} \sim N(0,\bV(s_{i-1},s_{i-1}\! +\!\Delta t_{i})),\; \qquad \label{eq:stTr}\\
 \text{(Phase correction)}\, &\hspace{-0.2cm} s_{i-1}\! =\! \argmin\limits_{s\in [0,\pi]} d\left(\frac{\bx_{i-1}}{\Omega},\bphi(s)\right),&\hspace{-0.1cm} \bkappa_{i-1} =  (\bx_{i-1} - \Omega\bphi_{i-1})/\sqrt{\Omega}  \label{eq:pc}
\end{eqnarray}

\noindent $\bC$ and $\bV$ are defined in Equations \ref{eq:Code} and \ref{eq:Code} respectively. 
The only difference between pcLNA and the standard LNA model is the phase correction step in equation \eqref{eq:pc}. 
There we find the phase-time, $s_{i-1}$, such that the point, $\bphi(s_{i-1})$, which is on the limit cycle, minimises the distance $d(\bx_{i-1}/\Omega,\bphi(s))$ to the state $\bx_{i-1}/\Omega$. If $d$ is the Euclidean distance metric, then $\bphi(s_{i-1})$ is the point on the limit cycle that is the closest to the state $\bx(t_{i-1})/\Omega$. It is also the intersection point of the limit cycle and the orthogonal transversal section at $\bphi(s_{i-1})$. 
The stochastic term is next adjusted to $\bkappa_{i-1}$ that lies on the transversal section, and is orthogonal to the tangential direction, which implies the elimination of noise in the unstable tangential direction. 
The search for the minimum distance in equation \eqref{eq:pc} can be performed by an optimization method (we use the standard Newton-Raphson method), and it is extremely fast, partly due to the search in the constraint space $[0,\pi]$. 
Note that a single set of solutions of the initial value problems in \cref{eq:ode,eq:Code,eq:Vode} of the LNA are used for computing the states at all time-points, which makes simulations extremely fast. 

As discussed earlier, we found that 3-4 phase corrections per cycle are needed to achieve high accuracy. 
In experimental settings, multiple time-points per cycle will be observed in order to study the oscillations. 
If more than 4 time-points are observed per cycle, one may perform the phase correction step only for some of time-points and perform standard LNA transitions for the rest time-points, which would speed up the computation slightly without seriously affecting accuracy.

\section{The pcLNA Kalman Filter}\label{seq:pcLNAKF}

We now consider time series observations of the system $ \by_0,\by_1,\dots,\by_N$ where $\by_i = \by(t_i)$ and $0\leq t_0 \leq \dots \leq t_N$. 
We assume that the observation $\by_i=(y_{1i},\dots,y_{qi})$ is related to the state $\bx_i=\bx(t_i)$ by the observation equation  
\begin{equation}\label{eq:obs1}
\by_i = \by(t_i) = \bB \bx_i + \bepsilon_i. 
\end{equation}

Here the matrix $\bB$ in  \eqref{eq:obs1} is a constant $q \times n$ matrix that determines how the observation relates to the underlying stochastic process. 
For example, $\bB$ might eliminate unobserved variables in $\bx$.
The $m$-dim vectors $\bepsilon_i$, $i=1,\dots,N$ are independent (between them and to $\bx$) measurement errors with normal distribution $ N(0,\bSigma_e)$.

We derive a Kalman Filter to compute the likelihood of model parameters $\btheta$
\begin{equation}\label{eq:likelihood}
    L(\btheta ; \mathbf{Y}) =  \prod_{i=0}^N P(\by_i | \bY_{(i-1)}; \btheta ), \quad \text{where }\bY_{(i-1)} = \{\by_{-1},\by_0,\dots,\by_{i-1}\}, \text{ for } i=0,1,\dots,N.
\end{equation} 
The Kalman Filter is a recursive algorithm (see \cite{KalmanFilter_book}) that uses the model's transition equations and Bayes rule to sequentially obtain the likelihood factors $P(\by_{i} | \bY_{(i-1)}; \btheta ) $ for $i=1,2,\dots,N-1$. We next provide the steps of the pcLNA Kalman Filter algorithm, and then discuss its details. 

 We use the notation $\by_{-1}=\{\bmu_0, \bSigma_0\}$ to denote the mean and variance matrix for the initial state, $\bx_0$. That is, 
\begin{equation}\label{eq:initialState}
\bx_0 = \Omega \bphi_0 + \sqrt{\Omega} \bxi_0  \sim N(\bmu_0, \bSigma_0). 
\end{equation} 
We also denote the p.d.f.\ of a MVN distribution with mean vector $\bom$ and variance-covariance matrix $\bS$ as $\cN(\cdot \, |\bom,\bS)$, the Euclidean distance between points $\bx$ and $\by$ in $\bR^n$ as $d(\bx,\by)$, while $\|\cdot \|$ denotes the Euclidean norm. 
The mean and variance of observation $\by_i$ are denoted by $\{\bmu^{(\by)}_i ,\bSigma^{(\by)}_i\}$, means and variances of priors $(\bx_i|\bY_{(i-1)})$ and $(\bkappa_i|\bY_{(i-1)})$ by $\{\bmu_i,\bSigma_i\}$ and $\{\bom_i,\bS_i\}$, respectively, with the symbol $^*$ denoting the parameters of corresponding posteriors $(\bx_{i-1}|\bY_{(i-1)})$ and $(\bkappa_{i-1}|\bY_{(i-1)})$.  

\begin{figure}
\centering
\includegraphics{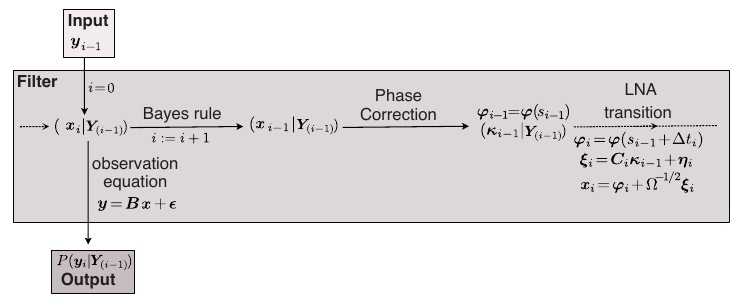}
\caption{\label{fig:KF}The iterative steps of the pcLNA Kalman filter. The filter starts by taking the inputs $\by_{i-1}$ and goes through a cycle of applying the Bayes rule, phase correction, and LNA transition to derive $(\bx_i|\mathbf{Y_{(i-1)}})$. 
Each time the filter reaches the latter step, it applies the observation equation to output a likelihood factor.}
\end{figure}

\begin{enumerate}[label=(KF\arabic*)]
\item \textbf{Inputs}: \label{KF:step1}
\begin{itemize}
\item Prior: $\by_{-1}=\{\bmu_0, \bSigma_0\}$ as in \eqref{eq:initialState}
\item Observed time series: $\{(t_0,\by_0),(t_1,\by_1),\dots,(t_N,\by_N)\}$, 
\item System size parameter $\Omega$,
\item $\bB$ and $\bSigma_{e}$ as in \eqref{eq:obs1}, 
\item $\bphi(t)$, $\bC(t)$, $\bV(t)$, $t\in [0,\pi]$ satisfying \eqref{eq:ode},\eqref{eq:Code}, and \eqref{eq:Vode}, respectively, 
\item $\bF(\bphi(t))$ as in \eqref{eq:ode}.
\end{itemize}
\item Compute $\bmu^{(\by)}_0 = \bB \bmu_0 $, $\bSigma^{(\by)}_0 = \bB\bSigma_0\bB^T + \bSigma_e$.\\[1em] \textbf{Output}: $P( \by_0| \by_{-1};\btheta) = \cN( \by_0 | \bmu^{(\by)}_0 ,\bSigma^{(\by)}_0) $. \label{KF:step2}\\
\item For iteration $i=1,2,\ldots,N-1$ \label{KF:step3}
\begin{enumerate}[label=(KF3\alph*)]
\item Compute the mean and variance of posterior $(\bx_{i-1}| \bY_{(i-1)})$ 
\begin{eqnarray*}
 \bmu^*_{i-1} &=& \bmu_{i-1} + \bSigma_{i-1} \bB^T (\bSigma^{(\by)}_{i-1})^{-1} \hat{\bepsilon}_{i-1}, \quad \hat{\bepsilon}_{i-1} = \by_{i-1} - \bmu^\by_{i-1}\\
 \bSigma^*_{i-1}&=& \bSigma_{i-1} - \bSigma_{i-1}\bB^T (\bSigma^{(\by)}_{i-1})^{-1} \bB\bSigma_{i-1}.
\end{eqnarray*} \label{KF:step3a}
\item Find $s_{i-1} = \argmin_{s\in[0,\pi]} d(\bmu^*_{i-1}/\Omega,\bphi(s))$. Then, set $\bphi_{i-1} := \bphi(s_{i-1})$, and compute the mean and variance of posterior $(\bkappa_{i-1}| \bY_{(i-1)})$ 
\begin{eqnarray*}
\bom^*_{i-1} &=& \Omega^{1/2}(\bmu^*_{i-1} - \Omega \bphi_{i-1}) \\
\bS^*_{i-1} &=& \Omega \bE_2 \left(\bSigma_{22,i-1}^* - \frac{ (\bSigma_{12,i-1}^*)^T \bSigma_{12,i-1}^*}{\bSigma_{11,i-1}^*} \right) \bE_2^T
\end{eqnarray*}
where
\[ \bSigma^*_{11,i-1} = \be_1^T\bSigma^*_{i-1} \be_1,\quad \bSigma^*_{12,i-1} = \be_1^T\bSigma^*_{i-1} \bE_2, \quad
 \bSigma^*_{22,i-1} = \bE_2^T\bSigma^*(t)\bE_2. \]\label{KF:step3b}
 Here, $\be_1 :=  \bF(\bphi_{i-1})/\|\bF(\bphi_{i-1})\|$, $\bE_2:=[\be_2 \cdots \be_n]$ with $\be_{j}^T \be^{}_{j}=1$, and $\be_{j}^T \be^{}_{j'}=0$, for $j\neq j'$,  and $j,j'\in \{1,2,\dots,n\}$.
\item Compute the mean and variance of prior $(\bx_{i} | \bY_{(i-1)})$ 
			\begin{eqnarray*}
				\bmu_i &=& \Omega \bphi_i + \sqrt{\Omega}\bC(s_{i-1},s_i)\bom^*_{i-1}, \, \bphi_i = \bphi(s_{i}),\, s_{i} = s_{i-1}+ \Delta t_i, \, \Delta t_i= t_i-t_{i-1} \\
				\bSigma_i &=& \Omega(\bC(s_{i-1},s_i)\bS^*_{i-1} \bC(s_{i-1},s_i)^T + \bV(s_{i-1},s_i)).
			\end{eqnarray*}
			\label{KF:step3c}
\item Compute $\bmu^{(\by)}_i = \bB \bmu_i $, $\bSigma^{(\by)}_i = \bSigma_i + \bSigma_e$.\\ \textbf{Output}: $P( \by_i|\bY_{(i-1)};\btheta) = \cN( \by_i | \bmu^{(\by)}_i ,\bSigma^{(\by)}_i) $. \label{KF:step3d}
\end{enumerate}
\end{enumerate}

The pcLNA Kalman Filter transitions are governed by the observation equation in \eqref{eq:obs1} and the pcLNA equations \eqref{eq:state}-\eqref{eq:pc}. As we illustrate in Figure \ref{fig:KF}, in steps \ref{KF:step2} and \ref{KF:step3d} we use the observation equation to derive the likelihood terms, while in step \ref{KF:step3a}, we use Bayes rule to derive the posterior distribution
$P(\bx_{i-1}| \bY_{(i-1)})$.
A Newton-Raphson algorithm can be used to find $s_{i-1}$ in the phase correction in \ref{KF:step3b}. 
Then, the posterior $(\bkappa_{i-1}| \bY_{(i-1)})$, is derived as a phase conditional distribution given that the noise component of the pcLNA ansantz in \eqref{eq:state}, after setting the phase to $s_0$, is orthogonal to the tangent $\bF(\bphi_{i-1})$, which ensures that the tangential direction noise is eliminated at the phase correction step. Details of the Kalman Filter steps are provided in Supplementary Information (SI) Section 1.

\subsubsection{Multiple time series} 
Typically data consist of multiple independent time series of observations, possibly with observations taken at different time-points. 
In this case, the joint likelihood of the multiple time series is simply the product of the likelihood of each time series that can be computed based on the above Kalman Filter. 
If the same initial condition is used for all time series, then the LNA system in \cref{eq:ode,eq:Code,eq:Vode} needs to be solved only once for all time series. 
This assumption might be suitable when the experimental design implies that the oscillating system is initialised in the same experimental conditions. 
For instance, for the \NFKB system that we consider, each time series correspond to different cells. 
These cells at the initial time-point are left at resting state for an appropriate time interval before applying an activating signal.
The noise around the initial condition is incorporated into the model by the distribution of $\bxi_0$. 
Therefore, the use of the same initial conditions is not only appropriate but also preferable because it allows us to draw inference for them from the data. 
Alternatively, the initial conditions might differ only by their phase in the oscillating cycle. Again the LNA system can be solved once and the phase of the system at the initial time-point can be identified by phase correction. 

Finally, if the initial conditions differ vastly one might wish to use a different initial condition for each time series with  additional computational cost. 

\section{System sensitivity analysis and identifiability}

Due to the coupling between the variables of non-linear dynamical systems in biology, it is often the case that parameters are fully or partially non-identifiable. 
Local sensitivity metrics such as Fisher information enable a study of parameter identifiability in the system and can therefore provide an \textit{a priori} indication of which parameter(s) are identifiable. 
Here, we will show how a decomposition of the Fisher Information matrix allows us to define parameter sensitivity coefficients that capture the effect of changes in parameter values on the likelihood.

Fisher Information quantifies the information that
an observable random variable carries about an
unknown parameter. 
If $P_{\btheta}(\bx)$
is the
probability density function of a continuous random vector $\bx$ depending on parameter vector
$\btheta=(\theta_1,\dots,\theta_K)$, the Fisher Information Matrix (FIM)
$\cI=\cI_{\btheta}$ is the $K\times K$ matrix with ($ij$)-th entry
\begin{equation}\label{eqn:FIM}
	\cI_{ij} = E\left[ \frac{\partial \ell}{\partial
	\theta_i}\frac{\partial \ell}{\partial \theta_j} \right]
	=-E \left[ \frac{\partial^2 \ell 
	(\btheta;\bx)}{\partial \theta_i \partial \theta_j} \right],
\end{equation}
where $\ell = \log P_{\btheta}$, and
$\theta_i$ and $\theta_j$ are 
the $i$th and $j$th components of the parameter vector $\btheta$. 
If $P_{\btheta}$ is Multivariate Normal (MVN) 
with mean vector $\bmu = \bmu(\btheta)$ and covariance matrix
$\bSigma =\bSigma(\btheta)$ then
\begin{equation}\label{eqn:fisher}
\cI_{ij}=
\frac{\partial\bmu }{\partial \theta_i}^T	\bSigma^{-1}
\frac{\partial\bmu }{\partial \theta_j}
+\frac{1}{2} \text{tr}
\left( 
	\bSigma^{-1}
	\frac{\partial\bSigma }{\partial \theta_i}
	\bSigma^{-1}
	\frac{\partial\bSigma }{\partial \theta_j}
	\right). 
\end{equation}
The FIM measures the sensitivity of $P_{\btheta}$
to a change in parameters in the sense that
\[
D_{KL}(P_{\btheta+\delta\btheta} ||
P_{\btheta} )
= \frac{1}{2}\delta\btheta^T\cI_{\btheta} \delta\btheta +\text{O}(||\delta\btheta ||^3)
\]
where 
$D_{KL}$ is the Kullback-Leibler divergence\footnote{$D_{KL}(P_{\btheta+\delta\btheta} ||
P_{\btheta} ) = \int P_{\btheta+\delta\btheta}(\bx) \log\left( \frac{P_{\btheta+\delta\btheta}(\bx) }{ P_{\btheta} (\bx)} \right) d\bx$, see \cite{CoverThomas}.} between the distribution of the random vector $\bx$ with parameter $\btheta$ and the distribution of $\bx$ with parameter $\btheta+\delta\btheta$. 
The significance of the FIM in
(\ref{eqn:FIM}) for sensitivity and
experimental design follows from its role as an approximation to the Hessian
of the log-likelihood function at a maximum.
Assuming non-degeneracy, the likelihood of a parameter value $\btheta$ near the maximum likelihood estimate $\btheta^*$ is
\begin{equation*} 
\ell (\btheta)
\approx \ell (\btheta^*) -\frac{1}{2} \sum_i \lambda_i^2
\left.\btheta^\prime_i\right.^2,
\end{equation*}
where
$\lambda_i^2$
are the eigenvalues ($\lambda_i$ the singular values), and 
$\btheta^\prime_i$ the coordinates of the parameter $\btheta$ with respect to an eigenbasis of the FIM in \eqref{eqn:FIM} and centred on $\theta^*$.
If we assume 
that the $\lambda_i$ are ordered so
that 
$\lambda_1^2 \geq \cdots \geq \lambda_K^2$
then it follows that near the 
maximum the likelihood is
most sensitive when $\theta^\prime_1$ 
is varied and
least sensitive when 
$\theta^\prime_K$ is.
Moreover, $\lambda_k$, $k=1,\dots,K$, is a 
measure of this sensitivity. 

Furthermore, because the FIM $\cI_\th$ is 
symmetric and positive semi-definite, 
it can be decomposed (e.g.\ by Cholesky decomposition) to 
$\cI = \Sy^T\Sy$
and the KL divergence  
\begin{equation}\label{eq:KLsens}
    \DKL (P_{\th+\dth} ||P_{\th})
 =  \frac{1}{2}\|\Sy\, \dth\|^2 +\text{O}(||\dth ||^3) = \frac{1}{2}\sum_{i,j,l} \delta \theta_j\delta \theta_l u_{ij}u_{il}  +\text{O}(||\dth ||^3) .
\end{equation}
The length, $\|\bu_k\|^2 = 2\DKL (P_{\th+ \delta \bm{e}_k} ||P_{\th}) + \text{O}(||\dth ||^3) $, of the $k$-th column vector $\bu_k = (u_{1k},\dots,u_{Kk})^T$ of $\bU$, measures the effects of a single unit change of the $k$-th parameter $\theta_k$ to the distribution $P_\th$, $k=1,\dots,K$, i.e.
  $  \DKL (P_{\th+\delta\bm{e}_j} ||P_{\th})
 =  \frac{1}{2} \delta^2 \| \bu_j \|^2 +\text{O}(|\delta|^3)$, where $\bm{e}_k$ the unit vector with only non-zero entry being the $k$-th entry.
It can therefore be used to study the sensitivity of $P_{\th}$ to changes in the parameter values.

If we consider the eigen-decomposition of $\cI$ and take the matrix $\bU$ above to have entries
\begin{equation}\label{eq:eigen}
u_{ik}=\lambda_i v_{ik}, 
\end{equation}
with $\bv_{i}=(v_{i1},\dots,v_{iK})$ the eigenvector corresponding to $\lambda_i$, then the row vectors, $\bu_{i}$, of $\bU$ have magnitude $\lambda_i$ indicating the overall sensitivity for changes of the parameter vector in the corresponding eigen-direction. 
The sign and magnitude of the entries of the eigenvectors reveal the correlations between parameters, much like principal component analysis.

\subsection{Computation of FIM for reaction network dynamics}\label{sec:compFIM}
The computation of the Fisher Information matrix requires the computation of the derivatives of the likelihood in \eqref{eqn:FIM}. 
If the data distribution is assumed to be Multivariate Normal then this computation reduces to computing derivatives of the mean vector and covariance matrix \eqref{eqn:fisher}. 
In the case where observations are time series with assumed model the continuous time-inhomogeneous Poisson process of reaction network dynamics, the computation of likelihood and even more the derivatives required for the FIM 
are very rarely computationally feasible. 
The LNA model gives Multivariate Normal distributions for time series observations and FIM computation is feasible, but it is often inaccurate for oscillatory dynamics as we explained in section 2. 
The pcLNA model requires a Kalman Filter to compute the likelihood of time series observations as a product of transition densities that are Multivariate Normal. 
It is possible but computationally intensive to calculate the required derivatives of these likelihood functions. 
Here we opt for computing the FIM with the likelihood function set as the joint probability distribution, derived in \cite{minasrand17} (see Supplementary information 1, section 9 for details), of trajectories transitioning between a large number of transversal sections of the limit cycle solutions of the studied systems under the LNA model. 
We ask whether the sensitivity coefficients of this FIM can predict which parameters are identifiable. 
Note that we use the pcLNA Kalman Filter described in section \eqref{seq:pcLNAKF} to compute the likelihood of time series observations. 
Both the pcLNA Kalman Filter likelihood function and the likelihood of trajectories transitioning between a large number of transversal sections are derived under the pcLNA model and they both control the instability of oscillatory dynamics in the tangential direction, only in a different way as explained in section 2.  
In section \ref{sec:resultss}, we will try to answer this question by applying our methodology to our exemplar systems and comparing the outputs with our sensitivity analysis predictions.

\section{Markov chain Monte Carlo algorithms}\label{sec:MCMC}
We implement a Bayesian analysis to conduct inference on the selected parameters within the model systems. Previous analyses of smaller systems \citep[e.g.][]{girolami11,burton2021} have shown that the choice of inference algorithm can severely impact the efficiency of the parameter estimation for even relatively small dynamical systems. The likelihood of time series observations is derived through the Kalman Filter and has the form of a product of conditional Multivariate Normal densities as explained above. 
It's not possible to derive the full conditional distributions for the parameters due to the complex dependence of the likelihood to the parameters and therefore we cannot utilise Gibbs samplers to conduct inference. 

Due to the nature of the dynamics of the systems, previous studies have found significant problems of poor mixing in chains when conducting analyses on small stochastic systems \citep[e.g.][]{fearnhead14}.
\gma{Taking also into account the presence of a large number of parameters that are strongly linked between each other and could present multi-modal probability distributions,} 
we implement 
a parallel-tempered Random Walk algorithm \citep{gupta18}.

\subsubsection{Parallel-tempered MCMC algorithm}

A feasible solution to attempt to circumvent the problem of local modes is trying to run a population of Markov chains in parallel, each with possibly different, but related stationary distributions. Information exchange between distinct chains enables the target chains to learn from past samples, improving the convergence to the target chain \citep{gupta18}.

Parallel tempering (PT) is an algorithm that attempts periodic swaps between multiple Markov chains running in parallel at different temperatures, where samplers with shallower energy landscape can traverse between multiple modes more efficiently. Each chain $j=1,...,J$ is equipped with an invariant distribution connected to an auxiliary variable, the temperature $\beta_j$, where $\beta_1=1$ followed by a sequence of decreasing positive parameters for $j=2,...,J$. These auxiliary variables scale the shallowness of the energy landscape, and hence define the probability of accepting an unsuitable move \citep{gupta18,hansmann}. An increase in the temperature, corresponding to a decrease in the auxiliary variable $\beta$, eases the traversal of the sample space. High temperature chains accept unfavourable moves with higher probability. As a result, higher temperature chains allow circumvention of local minima, improving both convergence and sampling efficiency \citep{chib}.

Defining the energy $E[\mathbf{\theta}]$ as the negative log-posterior evaluated at that parameter value, the Parallel Tempering algorithm for a parameter vector $\mathbf{\theta}$ can be illustrated as follows.
\begin{itemize}
    \item For $s=1,...,S$ swap attempts
    \begin{enumerate}
        \item[1.] For $j=1,...,J$ chains
        \begin{enumerate}
            \item[I.] For $t=1,...,N_{iter}$ iterations
            \begin{enumerate}
                \item[i] Propose a new parameter vector  $\theta^{'}(j)$ from a symmetric distribution $q(\theta^{'}(j)|\theta(j))$
                \item[ii] Calculate the energy $E_j=E[\theta^{'}(j)]$
                \item[iii] Set $\theta_{t+1}(j)=\theta^{'}(j)$ with probability $min(1, e^{-\beta_j \Delta E_j})$, where $\Delta E_j=E[\theta^{'}(j)]- E[\theta_{t-1}(j)].$ Otherwise, set $\theta_{t+1}(j)=\theta_t(j)$.
            \end{enumerate}
            \item[II.] Record the value of the parameters and the energy on the final iteration
        \end{enumerate}
        \item[2.] For each consecutive pair of chains (in decreasing order of temperature)
        \begin{itemize}
            \item[I.] Accept swaps with probability $min(1,e^{\Delta \beta \Delta E})$, where $\Delta E=E_j-E_{j-1}$ and $\Delta \beta= \beta_j-\beta_{j-1}.$
        \end{itemize}
    \end{enumerate}
\end{itemize}

If $\beta_j=1$, then the exact posterior distribution of interest is being sampled, hence, samples of the chain where $\beta_j=1$ are retained and summary statistics calculated, with other chains used for improving convergence and subsequently discarded. An additional step could combine use an importance sampling approach to combine the chains, thus further improving efficiency and avoiding discarding samples of higher temperature chains \citet{gramacy10}.

\section{Simulation studies}
\label{sec:resultss}

\subsection{Sensitivity analysis}

\begin{figure}
\centering
\includegraphics{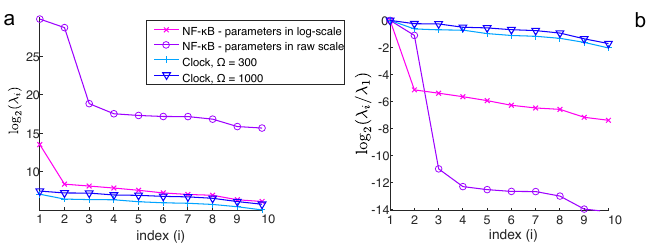}
\caption{\label{fig:sysanal}Singular values of the FIM for the pcLNA model of the two exemplar systems.
Plot (a) compares the 10 largest (see Supplementary Information Figure 1 for all) singular values of the FIM of: the \NFKB system with model parameters in log- and raw-scale, the Drosophila circadian clock model for two values of the system size $\Omega$ (see legend). 
Plot (b) compares the 10 largest normalised (divided by the largest singular value) singular values of the FIM of the same models as in (a) (see legend). The \NFKB system overall has larger sensitivities, but this is dominated by the sensitivity in few eigen-directions, while the circadian clock model is more balanced with a number of similarly sensitive eigen-directions.}
\end{figure}

System sensitivity analyses were performed for the two exemplar systems using Fisher Information around the values of parameters reported in the literature as broadly representing the system dynamics. 
The analyses, which are shown in Figure \ref{fig:sysanal}, examine both the absolute and relative sensitivity of the model through the eigenvalues of the FIM, but also the model sensitivity to changes in the values of each parameter. 

\begin{figure}
\centering
\includegraphics[scale=0.9]{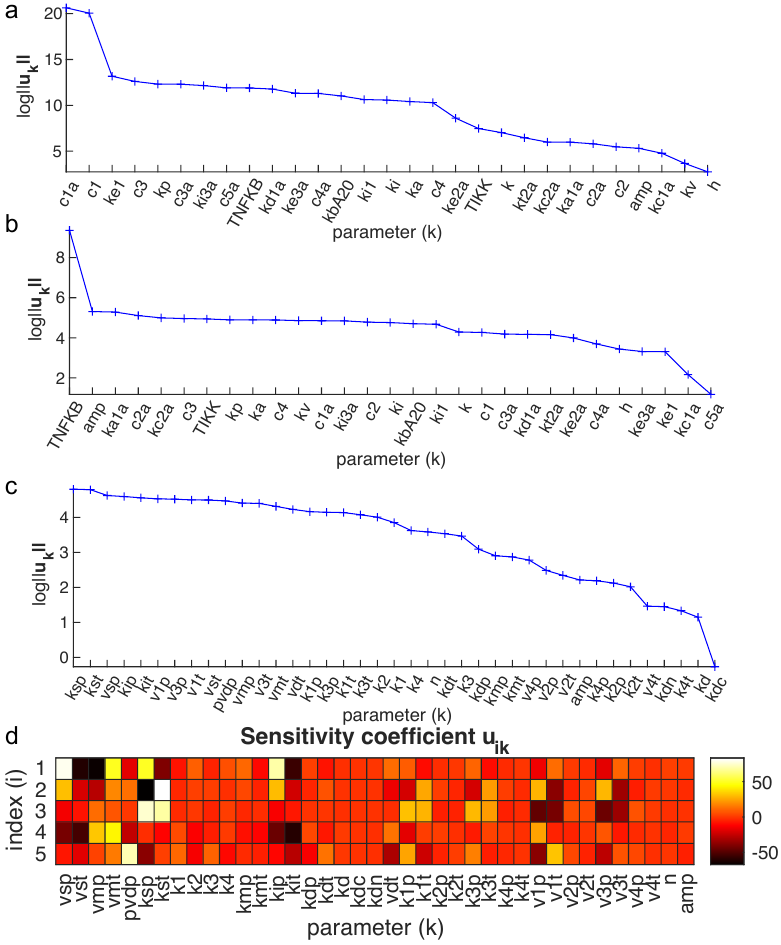}
\caption{\label{fig:sensAnalall}
Sensitivity coefficients for the pcLNA model of the \NFKB and the Drosophila circadian clock systems. 
Plots (a), (b) and (c) give the overall sensitivity coefficients, $\log(\| \bu_k\|)$ (as in \eqref{eq:KLsens}) for all parameters of the \NFKB system with parameters in raw scale (a), and log scale (b), and the circadian clock system with parameters in log-scale in (c). For the latter system, (d) presents a heatmap of the sensitivity coefficients $u_{ik}$ (see \eqref{eq:eigen}) corresponding to the top $5$ eigenvalues of the FIM.}
\end{figure}

The singular values of the FIM of the NF-$\kappa$B system with parameter changes in the log-scale are substantially larger than all other systems indicating a generally higher sensitivity in this model (see Figure \ref{fig:sysanal}a, and Supplementary Information (SI), Figure 1 
). 
However, the relative sensitivity of the same system is worse than any other considered system as the first two singular values dominate the magnitude of the FIM (see Figure \ref{fig:sysanal}b). 
The circadian clock model for both values of $\Omega$ exhibits a slow decrease in the first ten singular values of the system, which is translated as relatively small differences in the sensitivity of the model in a much larger number of directions of the parameter space, compared to the \NFKB system. 
This suggests that the circadian clock will provide better mixing of the MCMC chains since the changes in the likelihood over the parameter space are more smooth compared to the \NFKB system.
In contrast, the sharp decrease of the singular values of the \NFKB system indicate that the likelihood is much more sensitive to changes in a small number of directions of the parameter space (one for the log-scale and two for the raw scale) compared to any other direction. When conducting inference, the choice of scale to conduct inference on will also change depending on this variation in sensitivity.

From the sensitivity analysis, it is apparent that the two parameters associated with the two largest singular values of the \NFKB system in the raw scale were those corresponding to the mRNA synthesis rates for the variables I$\kappa$B$\alpha$ and A20, namely $c1$ and $c1a$ respectively. Notice the 
corresponding high values of the overall sensitivity $\|\bu_k\|$ in Figure \ref{fig:sensAnalall}a. 
Different parameters were highlighted as sensitive on the log-scale, with substantially higher sensitivity coefficients primarily for total \NFKB concentration (TNFKB) compared to all other parameters (see Figure \ref{fig:sensAnalall}b).

The slow decay in the singular values of the \emph{Drosophila} circadian clock is reflected also in the parameter sensitivity coefficients (see Figure \ref{fig:sensAnalall}c) with a larger number of parameters having closer sensitivities compared to those in the NF-$\kappa$B model. 
In particular, the sensitivities coefficients suggest that more parameters 
could be estimable. 
Figure \ref{fig:sensAnalall}d, reveals relations between parameters, which might be reflected in the posterior distributions.

We used these sensitivity analyses to guide the
choice of parameters to estimate within the Bayesian inferential framework. 
In particular, we highlight the differences between estimating parameters with high sensitivity coefficients compared to estimating those with low sensitivity coefficients.

We see whether there is a clear dependence between the sensitivity coefficients and the precision of posterior distributions. 
We also consider scenarios where a mix up of  parameters with high and low sensitivity and coefficients is estimated.

\subsection{Parameter estimation\label{sec:parest}}

\gma{Here we consider a number of examples of applying our method for parameter estimation. 
We aim to demonstrate the use of our method, but also to investigate the relation between the parameter sensitivities and the implementation and inference of Bayesian computational methods.
}

For each of the different settings, ten independent stochastic trajectories were generated using either the pcLNA simulation algorithm \citep{minasrand17} or the full SSA, with parameters fixed at those from available literature \citep{ashall,drosmodel}.  
This is a modest sample size for typical experiments of those systems.
\gma{In all of the following examples, we derive posterior estimates for a selection of parameters. 
The parameters not included in the inference are fixed at the respective values from the literature used to simulate data.

We use the parallel-tempered algorithm described in section \ref{sec:MCMC} with $N_{iter}=50$ iterations between swaps and $S=600$ swap attempts. 
The proposal distributions $q$ are Multivariate Gaussian centred on the current parameter value.
The covariance matrix of the proposal distribution has a diagonal form (except when stated otherwise). 
The covariance matrix is adaptively updated for the first cycles of each run. 
That is, before each of the first $S_A=200$ swap attempts, we compute the acceptance rate of the last $N_{iter}$ iterations and if the acceptance rate is below $0.2$ or above $0.3$, we multiply the current proposal covariance matrix by $1-e$ or $1+e$ ($e=0.2$), respectively. Before the $(S_A+1)$-th swap attempt, we set the proposal variance to the average proposal variance in the last $50$ cycles, and stop updating it.

Prior distributions were specified on each of the rate parameters on the raw scale, whilst inference was conducted both on the raw scale and on the log scale with appropriate Jacobian term applied to the acceptance probabilities in the case of the latter. 
Alternatively, priors could be specified on the log-scale, removing the need for the Jacobian term. Relatively vague Gamma(1,10) priors were specified on rate parameters and inverse Gamma IG(0.001,0.001) for the error variance. 
Unless stated otherwise, we assume $\bB$ in \eqref{eq:obs1} is an identity matrix, that is all species are measured at all time points.
Posterior distributions are computed using the chains with temperature parameter $\beta_j=1$ and after dropping the first $10000$ iterations.} 

\subsubsection{NF-$\kappa$B system}

We first run the 
\gma{parallel-tempered} Markov chain Monte Carlo (MCMC) described \gma{above} 
to the NF-$\kappa$B system with parameters in raw-scale.
We should note that it is generally preferable to run the sensitivity analysis and estimation in log-scale. 
This is because the raw-scale is affected by the measurements units used to set the parameter values. 
A change in the measurement units can largely influence the sensitivities. 
In this first example, we simply consider the raw-scale because it provides an informative illustration of applying our methods.

The inference algorithm was initially run for the NF-$\kappa$B system with two unknown parameters ($c1a$ and $c1$) identified from the sensitivity analysis, with other parameters fixed at values from the literature. Posterior mean estimates were 1.57e$^{-7}$ (95\% credible interval (1.08e$^{-8}$, 3.08e$^{-7}$)) for $c1a$ and 1.16e$^{-7}$ (95\% credible interval (1.50e$^{-8}$, 2.05e$^{-7}$)) for parameter $c1$. The ground truth values were these were both $1.4e^{-7}$. 

\begin{figure}
\centering
\includegraphics[scale=0.3]{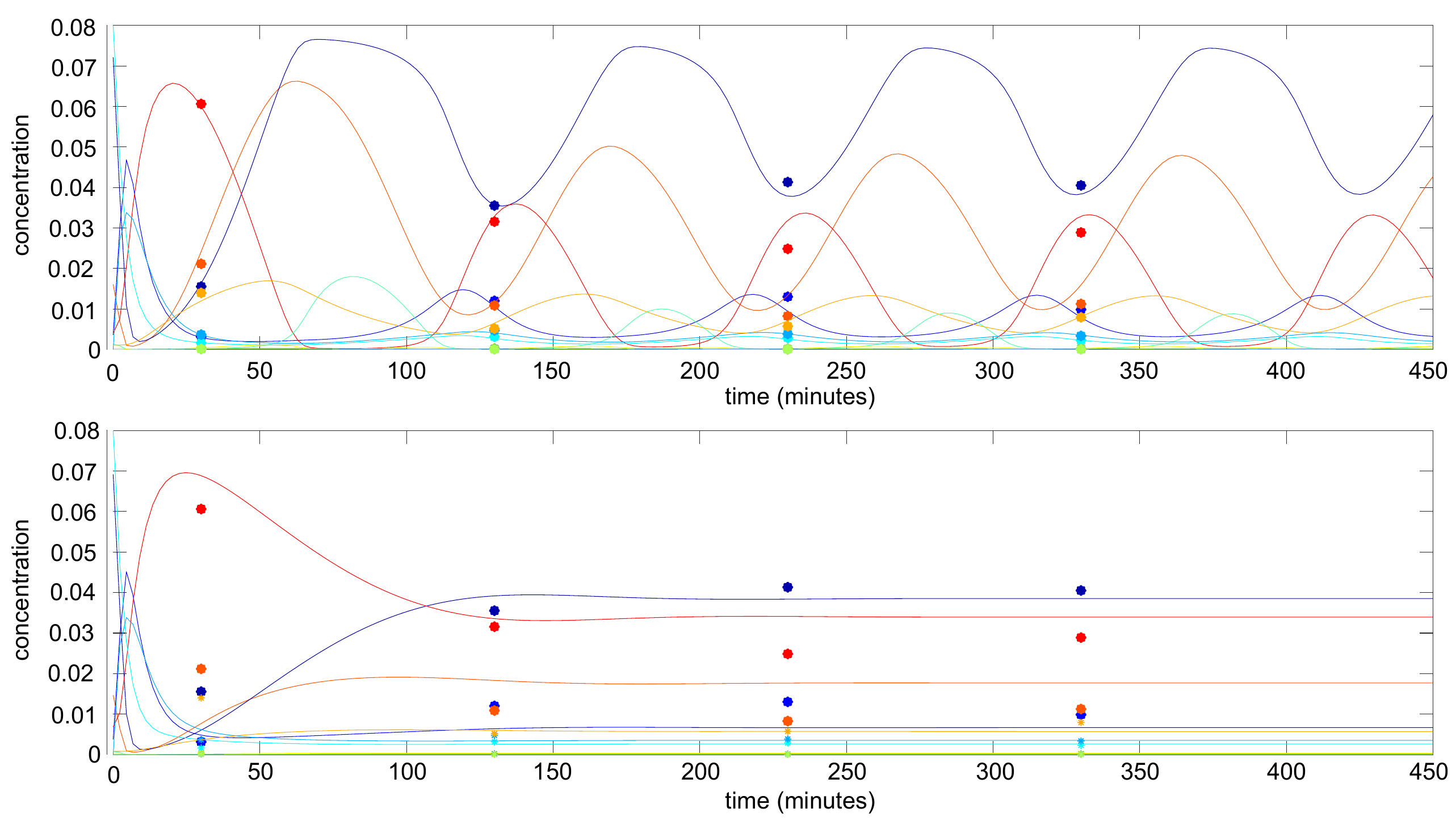}
\caption{\label{fig:postsimpath}Deterministic solutions of the NF-$\kappa$B ODE system evaluated at the two posterior modes highlighted by the parallel tempering algorithm. The upper plot corresponds to a value close to the correct values corresponding to the major mode, whilst the lower plot corresponds to values close to zero obtained in the minor mode, giving no oscillations beyond the first peak. The overlayed points correspond to the value of a randomly selected data sample used in the inference. The colors represent the variables of the system.}
\end{figure}

Of interest in these posteriors was a second minor mode close to zero. This second mode 
was studied further to determine possible reasons that the tempered algorithm visited this mode and the biological implications of this. Figure \ref{fig:postsimpath} shows the deterministic solution to the system with parameters $c1a$ and $c1$ at the values from the literature and also again when replaced with the mean of this lower mode. This shows a clear problem with the discrete time-points used for inference, as the observation times traditionally measured correspond only to the peaks of the nuclear NF-$\kappa$B. In this case, the algorithm correctly fails to reject parameters that give deterministic model solutions without oscillations as the level of nuclear NF-$\kappa$B remains constant at a similar level to the peak height in the oscillatory system. Biologically, this corresponds to a system with zero feedback between the corresponding variables.

As the parallel tempered algorithm was the only algorithm to effectively explore the multimodal parameter space (see SI, Figure 2
), all further analyses were conducted using this algorithm. We further varied the parameter settings in the following ways.

\subsubsection{Log-scale inference}

The sensitivity analysis also showed variability depending on whether the parameters were transformed or not prior to conducting analysis. 
Whilst parameters $c1a$ and $c1$ were the most sensitive on the raw scale, when transforming parameters to a log scale, other parameters had higher sensitivity coefficients. 
There is one dominant singular value  in the system (see Figure \ref{fig:sysanal}) 
and  
the total concentration of NF-$\kappa$B (TNFKB) was the dominant parameter, with additional lower sensitivities to parameters (TNF$\alpha$) dose and $kbA20$. 
Additional parameters with much lower sensitivities, namely $kc1a$ and $c5a$, were included to test how these would behave in the inference. 
The total NF-$\kappa$B parameter determines the overall levels of all variables in the system.
Prior distributions were kept similar to those in the previous analysis, specified on the un-transformed scale, but an appropriate Jacobian term was introduced into the Metropolis acceptance step to ensure the correct invariant distribution was reached. 
In the case of the log transform such that $\psi_i=\log(\theta_i)$, with prior specified on the raw scale, the log posterior is
$\log(\pi({\bm \psi}|\bY))= \log(L({\bm \psi}; \bY))+\sum_i\log(p(\exp(\psi_i)))+\sum_i \psi_i + constant$, where $p(\cdot)$ are the priors on the raw parameters, in this case Gamma distributions.

In each case, four chains were again run in parallel, with 20000 iterations consisting of 400 swaps in total, with swaps proposed every 50 iterations. The $\beta_j$ parameters across four chains were set at $(1,0.5,0.3,0.1)$. 
\begin{table*}[h!]
{%
\begin{tabular}{ccccc}
\\
\hline
Parameter& True value & Posterior mean & \multicolumn{2}{c}{95\% posterior CI}  \\ 
  \hline    
    TNFKB  &     0.08   &   0.079942     & 0.079432  &    0.080404\\
    kc1a   &   0.074   &  0.1012  &    0.037719   &    0.16624 \\
    c5a&2.2$\times 10^{-5}$&1.55$\times 10^{-4}$&  4.1757$\times 10^{-5}$ &    2.6$\times 10^{-4}$\\
    kbA20&0.0018 &    0.0014773 &     0.0011819 &     0.0017734\\
    dose&      1&   0.94905  &     0.83732   &     1.0651\\
   \hline
\end{tabular}}
\caption{\label{tab:lognfkb}Posterior estimates for the NF-$\kappa$B system with parameter inference conducted on the log-scale and all 11 variables measured. Data simulated from the pcLNA algorithm.}
\end{table*}
Table \ref{tab:lognfkb} presents posterior summaries of the results, in which the parameters can be reliably estimated. Posterior means were relatively close to the true parameter values. 
The credible interval of TNFKB is much narrower than those of all the other parameters.

\subsubsection{Unobserved variables}

Further analyses were run under the setting that measurements of some of the biological species are not easily obtained experimentally. In this case, the matrix $\mathbf{B}$ in the Kalman filter is no longer an identity matrix, but contains some zero elements along the diagonal also. Table \ref{tab:lognfkb4var} shows the results for the same set-up as Table \ref{tab:lognfkb} but assuming only nuclear and cytoplasmic NF-$\kappa$B, cytoplasmic I$\kappa$B$\alpha$ and A20 are measurable (i.e. the concentrations of the molecular species that are in protein form). 

\begin{figure}
\centering
\includegraphics[scale=1]{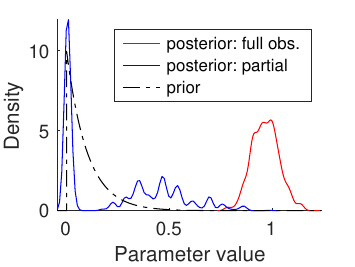}
\caption{\label{fig:unobspriorpost}Prior distribution (dashed line) vs posterior marginal density (solid line) for the dose parameter when only four of the eleven states are measured. Estimates correspond to those in Table \ref{tab:lognfkb4var}.}
\end{figure}

The parallel tempering algorithm converged similarly in this case, but the ability to estimate some of the parameters deteriorated. Specifically, the dose parameter posterior distribution was very similar to the prior distribution, suggesting the data is non longer informative on this parameter (Figure \ref{fig:unobspriorpost}) and the estimate of parameter $kc1a$ was negatively biased towards the prior mean.
The other parameters do not appear greatly impacted by the reduced measurement matrix. 
As the sensitivity coefficients are a sum over the sensitivities across all state variables in the system, if the dominant variable showing sensitivity to a given parameter is no longer measured, then the ability to practically identify that parameter will be removed. The number of dominant principal directions may also change.

\begin{table*}
{%
\begin{tabular}{ccccc}
\\
\hline
Parameter& True value & Posterior mean & \multicolumn{2}{c}{95\% posterior CI}  \\ 
  \hline
   TNFKB&   0.08&  0.091465&  0.061862&   0.12655\\
    kc1a&  0.074&  0.070118& 0.0030725&    0.2748\\
    c5a&2.2e-05&0.00046874&2.5228e-05&0.00089487\\
    kbA20& 0.0018&  0.049507&0.00040189&   0.26541\\
    dose&      1&   0.23781&0.00015067&   0.75594 \\
  \hline
\end{tabular}}
\caption{\label{tab:lognfkb4var}Posterior estimates for the NF$\kappa$B system with parameter inference conducted on the (natural) $\log$-scale and only four variables measured.  Data simulated from the pcLNA algorithm using the true value.}
\end{table*}

\begin{table*}
{%
\begin{tabular}{ccccc}
\\
\hline
Parameter & True value & Posterior mean & \multicolumn{2}{c}{95\% posterior CI} \\ 
  \hline
    vsp&1&  1.0127&   1.0029&   1.0227\\
    vst&1& 0.99898&  0.99017&   1.0077\\
    vmp &  0.7& 0.71984&  0.71099&   0.7298\\
    vmt  & 0.7& 0.70609&  0.69656&  0.71541\\
    pvdp& 2&    1.98&    1.962&   1.9963\\
    ksp &  0.9& 0.89508&  0.88843&  0.90132\\
    kst  & 0.9& 0.90119&  0.89581&   0.9058\\
    kip&1&  1.0134&   1.0047&   1.0223\\
    kit&1&  1.0019&  0.99284&   1.0118\\
    sigma & 0 & 3.298$\times 10^{-5}$  &  2.5738$\times 10^{-5}$  & 4.1863 $\times 10^{-5}$ \\
   \hline
\end{tabular}}
\caption{\label{tab:dros1000}Posterior summary statistics for the estimated parameters in the Drosophila circadian clock with $\Omega = 1000$.  Data simulated from the SSA and inference conducted on a (natural) log-scale.}
\end{table*}

\subsubsection{Drosophila circadian clock}

Analyses were also conducted using the Drosophila circadian clock system with similar settings to those in the \NFKB system. 
As noted previously, the clock system (considered in the log-scale of parameters) has higher sensitivities to a larger number of parameters and hence here nine parameters were firstly estimated. 
For the Drosophila clock, data were simulated directly from the Gillespe algorithm (SSA) \citep{gillespe} to confirm the correct stochastic trajectories can still be obtained when fitting to the exact algorithm.
Specifically we used $10$ SSA simulated trajectories and obtained observations every $8$ hours for all variables. The data are shown in Figure \ref{fig:detsols}. 
 
It would be expected that results under this setting could perform worse due to any potential discrepancies between the approximation of the stochastic dynamics using the pcLNA and the exact stochastic algorithm. 
Table \ref{tab:dros1000} shows posterior summary statistics. 
All estimates were very close to the true values, with credible interval coverage being high. Five out of nine parameters have the true value lying in the $95\%$ posterior Credible Intervals and the rest are very close.
The estimation of an additional observational error may explain some of the small discrepancies in parameter estimates, although the magnitude is negligible. 
Figure 4
in Supplementary Information (SI), compares the posterior to prior densities clearly demonstrating the gain of information using the data.

\begin{table*}[h!]
\begin{tabular}{ccccc}
\hline
Parameter & True value & Posterior mean & \multicolumn{2}{c}{95\% posterior CI} \\ 
  \hline
    vsp&      1&1.0062 &    0.97569&1.0389\\
    vst&      1&1.0565&1.0135&1.0912\\
    vmp&    0.7  &   0.78127 &     0.74592  &   0.83004\\
    vmt&    0.7& 0.86162& 0.80802& 0.90271\\
    pvdp&     2&2.0707& 2.007&2.2463\\
    ksp&0.9& 0.92547& 0.89496& 0.98938\\
    kst&0.9& 0.91072& 0.89053& 0.93883\\
    kip&  1&1.1758&1.1386&1.2432\\
    k&     1&1.2019& 1.153& 1.258\\
    sigma &  0  &  0.011867&0.010275&0.013883 \\ 
   \hline
\end{tabular}
\caption{\label{tab:dros300}Posterior summary statistics for the estimated parameters in the Drosophila circadian clock with $\Omega = 300$.  Data simulated from the SSA and inference conducted on a (natural) log-scale.}
\end{table*}

\subsubsection{Varying stochasticity}

The scale of stochasticity in the system is controlled by the system size $\Omega$. 
While in the example above, the fairly low value of $\Omega=1000$ is used, we here consider an even smaller value, namely $\Omega=300$, with same observed time-points, variables, and sample size as above. 
Reasonable estimates of the parameters were still obtained (Table \ref{tab:dros300}) but slightly higher bias in the posterior estimates and credible intervals were observed. Three out of nine parameters have the true value lying in the $95\%$ posterior credible intervals and the rest are fairly close, less so than when $\Omega=1000$. 
The observation error variance was also inflated in this case (posterior mean $0.011867$ compared to $0.00003$ in Table \ref{tab:dros1000}.).

\subsection{Parameter sensitivities and estimation}

Providing the best algorithm for implementing our method is beyond our scope here, but we present some practical guidance based on our experience with inference in these specific models. Unsurprisingly, the implementation that seemed to affect the mixing and convergence of chains significantly was the proposal covariance matrix of the proposal. Relative changes in effective sample size (ESS) were monitored due to the fact that ESS is unlikely to be an ideal absolute metric of efficiency in multimodal distributions \citep{vehtari21}.

In other experiments, we also tested the use of the Fisher information matrix within the proposal mechanism. The use of adaptive sampling in the burn-in phase automated the choice of proposal distribution and greatly alleviated these problems. 
That is, we set the proposal covariance matrix equal to the inverse of a Fisher Information multiplied by an adaptively-adjusted constant, with the adaptive part conducted as above.
The Fisher Information matrix was computed using the true parameter values used to simulate the data, and therefore the inverse Fisher Information matrix provides (under some conditions) an asymptotic approximation of the covariance matrix at the true parameter value. 
However, this simulation scenario is unrealistic because (at least) some of the parameter values are in practice unknown. 
The inverse Fisher information can also be far from the optimal choice for the proposal when its value at the current parameter vector substantially differs from that at the true parameter vector. 

We next wish to establish whether the sensitivity analysis performed earlier provides information that is useful for parameter estimation. 
We use the sensitivity coefficients $\|\bu_k\|$ (see \eqref{eq:KLsens} and Figure \ref{fig:sensAnalall}), as an overall measure of the sensitivity of the $k$-th parameter, $k=1,2,\dots,K$. 
We perform various runs of the parallel tempered algorithm. In each run, we select a number of parameters to be estimated based on their sensitivity coefficients, while we keep the rest fixed at their values used for simulation.
The runs we consider either estimate only the highest sensitivity parameters, or the lowest sensitivity parameters or a mix. 

To study the relation of the sensitivity coefficients $\|\bu_k\|$ with posterior variability, we compute the log-posterior Coefficient of Variation (CoV, posterior standard deviation/posterior mean) of each parameter chain, transformed back in raw scale.  
Note that the prior CoV is $1$ for the priors used here. 

\begin{figure}[h]
\centering
\includegraphics[scale=1]{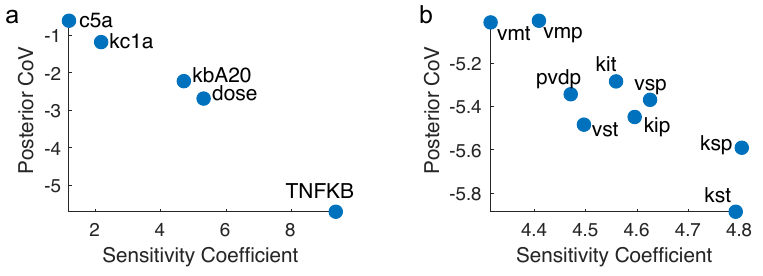}
\caption{\label{fig:sensvssd}Sensitivity coefficients plotted against posterior coefficient of variation (CoV) in the NF-$\kappa$B system (a) and Drosophila circadian clock (b). Both the posterior CoV and the sensitivity coefficients (i.e.\ the lengths $\| \bu_k\|$ in \eqref{eq:KLsens})) are in (natural) log-scale.}
\end{figure}

Figure \ref{fig:sensvssd} considers the case where a mix of high and low sensitivity parameters of the NF-$\kappa$B and circadian clock system are estimated. 
A power law relation with a negative exponent is observed in both of these cases. 
The TNFKB parameter has a much smaller posterior CoV than all other parameters. 
The circadian clock system shows less clear distinction across parameters, as all parameters estimated have more similar sensitivities than in the \NFKB system. 

\begin{figure}
\centering
\includegraphics[width=\textwidth ]{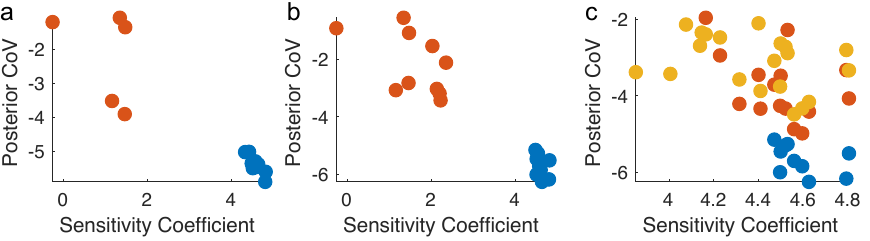}
\caption{\label{fig:CovAll}Posterior Coefficients of Variation (CoV) against the overall sensitivity coefficients $\|\bu_k\|$ (see \eqref{eq:KLsens}) for different runs of PT MCMC. 
The results in (a) are for a run with the 5 parameters of highest (blue circles) and lowest (red circles) $\|\bu_k\|$ are estimated, in (b) 10 parameters of highest (blue circles) and lowest (red circles) $\|\bu_k\|$ are estimated, and (c) 10 (blue circles), 15 (red circles) and 20 (yellow circles) parameters of highest $\|\bu_k\|$ among all parameters. In each run, the rest of the parameters are fixed to their true value. Both the CoVs and the sensitivity coefficients are (natural) log-transformed.}
\end{figure}

We next run the PT MCMC algorithm 
first estimating 5 parameters with the highest sensitivities, and then the 5 parameters with the lowest sensitivities in the circadian clock system. 
In Figure \ref{fig:CovAll}a, we see a clear distinction between the values of the posterior CoV in the two runs, with much smaller CoV for the high sensitivities. 
This result persists when we add more parameters with distinct sensitivities (see Figure \ref{fig:CovAll}b). 
The power law relation with negative exponent between the posterior CoV and the sensitivity coefficients is present when we estimate the 10 highest sensitivity parameters and then add the next 5 and 10 parameters in the sensitivity ranking (see Figure \ref{fig:CovAll}c). Clearly, high sensitivity parameters are estimated with higher posterior precision, despite similar prior CoVs. 

\begin{figure}
\centering
\includegraphics[width=\textwidth]{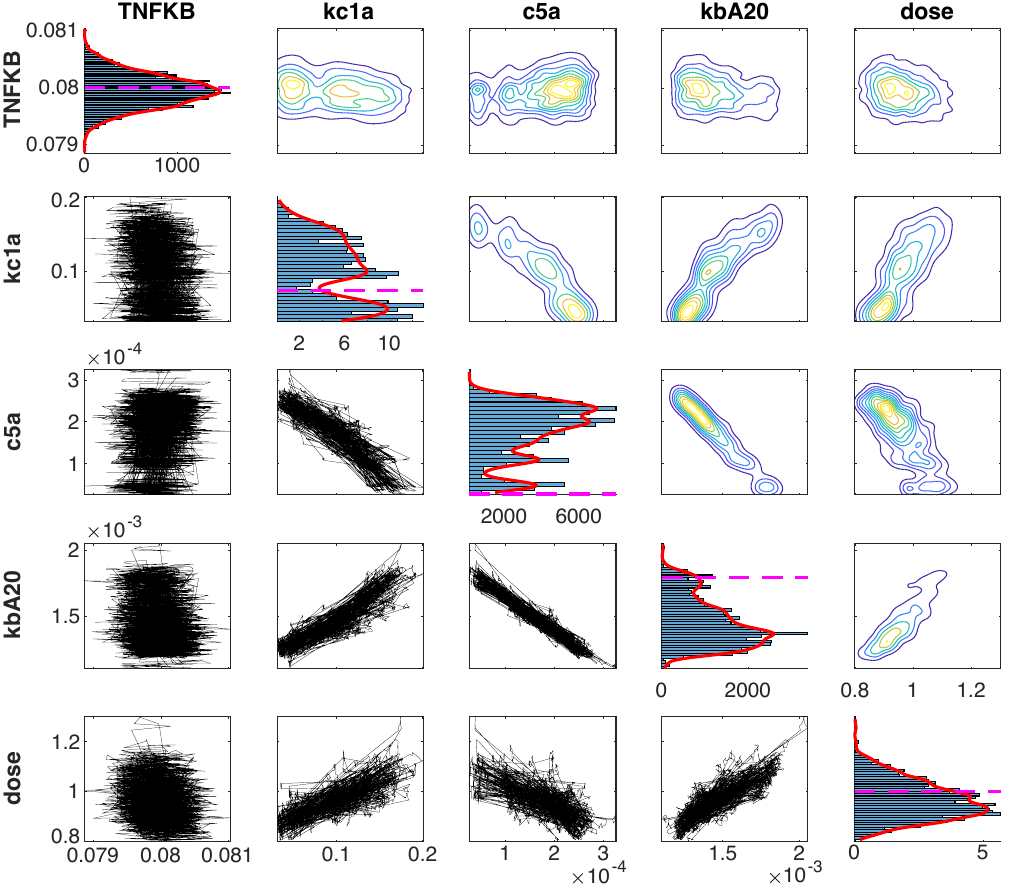}
\caption{\label{fig:2dchains}Posterior inference for the \NFKB system estimating 5 parameters (see plot titles). The histograms, smoothed densities and true values of each parameter (main diagonal), and the chains (lower diagonal) and contour plots of posterior distribution (upper diagonal) of parameters pairs are presented.}
\end{figure}

The mixing of the chains of high sensitivity parameters tends to be much better than for low sensitivity parameters.
This is observed in Figure \ref{fig:2dchains} and SI Figure \ref{nfkbchains_fig10}, and also reflected into the Effective Sample Size (ESS) of parameters. 
For instance, the TNFKB parameter that has much larger sensitivity coefficient than any other parameter in the system has ESS $1732.7$ compared to ESS around $438$ for the other estimated parameters. 
The low posterior correlation of TNFKB with the rest of the estimated parameters that are highly (positive or negatively) correlated seems to affect mixing (see also SI Figure \ref{low5_chains}). 

The improved mixing for high sensitivity compared to low sensitivity parameters is also observed in the runs on the Drosophila circadian clock system. 
This is also generally reflected to the ESS values (see SI Figures \ref{high5_chains}, \ref{low5_chains}). 
For parallel tempering, the mixing depends on the number and the values of temperatures used. 
We saw that increasing the number of temperatures close to one improves mixing (see SI tables \ref{sitab:essTab1}, \ref{sitab:essTab1}). 
We also know that mixing depends on the proposal variance. 
The mixing can improve by selecting a more appropriate proposal variance. 
To explore this further, we run the PT MCMC algorithm to estimate only two low sensitivity parameters ($v4t$ and $k4t$), which are highly correlated in the likelihood. 
This allows the adaptive sampler to select the proposal variance more freely than when sampling more parameters, because adaptation only depends on the acceptance rate for those parameters. 
The result is that mixing is similar with ESS slightly decreased, but the posterior variance is much larger, and the two marginal posterior distributions are much closer to the prior distribution (see SI Figures \ref{k4t_chains}, \ref{k4t_priorVpost}), compared to the case of estimating five parameters including these two.
On contrary, the results are very consistent when we restrict estimation to two high, instead of two low, sensitivity parameters (see SI Figure \ref{high2_priorVpost}). 
If one ignores the sensitivity analysis for these parameters, it can easily misinterpret the results in many ways.

\subsection{Comparison to restarting LNA of \citet{fearnhead14}}\label{sec:restartLNA}

We compare our approach to that of \citet{fearnhead14} (henceforth referred to as restarting LNA), which also uses the LNA as the underlying model. The difference is that the restarting LNA Kalman Filter replaces the phase correction implemented in our method with a resetting of the initial condition of the ODE in \eqref{eq:ode} that sets the stochastic component $\bxi_{t_i}$ to zero before transition from time $t_{i}$ to $t_{i+1}$. 
The restarting LNA makes no assumptions on the dynamics of the LNA, and therefore could potentially be applied to systems of any dynamics. 
The rough idea is to make LNA steps short transitions by frequently resetting of initial conditions; since LNA is accurate in short transitions this should improve its accuracy.  
The resetting of initial conditions of the ODE implies that the solutions of all the ODEs involved in the LNA are solved between each resetting times. 
The trajectory of the restarting LNA follows (in some sense) the (dynamic) posterior mean, even when this is rapidly changing, and it's therefore smooth only between transitions.
This is unlike the pcLNA that, which as we described in section \ref{sec:pcLNA}, for a given parameter vector, solves the system once for all the observed time interval, and adjusts only the time-phase of the ODE solutions. 

We implemented the same PT MCMC used above with the likelihood now computed under the restarting LNA described in  \citet{fearnhead14}, but all the remaining steps unchanged. 
We saw that the posterior distributions are similar to the pcLNA, with main difference the speed of implementation, which is about 2 times slower. 
The main factor for this slow down is the extra calls of the ODE solvers.
This is despite that ODE solutions are obtained for the same time-interval.

\begin{figure}
\centering
\includegraphics[scale=1.125]{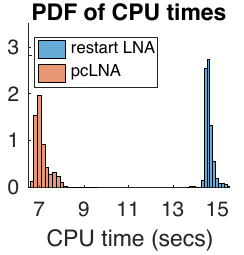}
\caption{\label{fig:histRestartLNA}Histograms of the CPU times of one iteration in seconds for pcLNA and restarting LNA (see legend). The increase in CPU time with restarting LNA is roughly 2 fold.}
\end{figure}

\section{Discussion}

The proposed approach of combining pcLNA with parallel tempering was highly successful in retrieving accurate estimates of parameters from simulated data in large stochastic non-linear dynamical systems with oscillations, conditional on the level of sensitivity of the model output to those parameters being sufficiently high. The sensitivity analysis was able to predict which parameters were estimable from observed data, across the tested simulation settings (different variables, varying Omega and Gillespe and pcLNA simulated data). These highly-parameterised systems have not previously been estimable due to the computational demand of calculating the conditional distributions of the state process at the desired timepoints. Our simulations have highlighted a variety of challenges of the systems, including (non-)identifiability, data resolution and multimodal distributions common to these types of systems. 

We emphasise the importance of conducting sensitivity analysis on non-linear dynamical systems prior to conducting inference. Attempting to conduct inference in partially- or non-identifiable systems risks erroneous results, even for parameters that should be identifiable. 
In the case of the parallel tempering algorithm, 
some of the issues observed were high posterior variability, 
posterior marginals being similar to the priors specified on them, and poor mixing. With too many non-identifiable parameters the problems deteriorate. Our results (see Section \ref{sec:parest}) also show the importance of considering the discrete time points that are used to conduct inference. In some instances these may be determined by the experimental set-up, but as modern biotechnologies continue to advance, the choice of resolution and interrogation of the underlying continuous process will be increasingly important. Sensitivity analyses or inference should also allow for varying temporal resolutions in the data \citep[e.g.]{baggenstoss}, particularly in the presence of observation errors.

Our approach for Bayesian inference using the tempered algorithm was able to account for bifurcations in parameter space and also highlight alternative parameter combinations consistent with the data that warranted further study. \citet{girolami11} highlight that extending the manifold MALA algorithm to a manifold Hamiltonian Monte Carlo (HMC) algorithm facilitates implementation within a parallel tempered algorithm and this may further improve the results found here. In particular, the improved convergence rate of parallel tempering combined with the improved local mixing of the geometric algorithms. It should be noted, however, that this does require significant increased computational power in the calculation of the derivatives of the metric tensor (i.e. FIM) and improvements in efficiency could be hindered by this. 

We have shown that the ability to estimate parameters accurately and precisely varies along a variety of axes within these stochastic models. Increasing number of large eigenvalues of the mechanistic model is shown empirically to be proportional to the number of estimable parameters in the model and the loadings of the principal eigenvalues are highly indicative of which parameters are estimable. Similarly, increasing stochasticity unsurprisingly hindered the ability to obtain more precise parameter estimates via uncertainty in the corresponding posterior marginals. In many experimental designs, it may not be feasible to measure all of the biological species and our results show this can hinder the estimation of some parameters, but others are still estimable if the sensitive outputs are still measured.

Our approach has made significant advances in the ability to conduct exact and fast Bayesian inference in high-dimensional stochastic systems of reaction networks, whilst being similarly applicable to other stochastic models based on stochastic differential equations, such as those found in epidemiology and population dynamics. Remaining challenges exist in improving optimal experimental design, as well as developing optimum block updating strategies for the parameters when single parameters may dominate.

\section*{Acknowledgement}
BS acknowledges funding from BBSRC grant BB/K003097/1 and the Edinburgh Mathematical Society.

\bibliographystyle{bafiles/ba}
\bibliography{mainForMinorRevisionSubmission}

\begin{thebibliography}{56}
\newcommand{\enquote}[1]{``#1''}
\expandafter\ifx\csname natexlab\endcsname\relax\def\natexlab#1{#1}\fi
\expandafter\ifx\csname url\endcsname\relax
  \def\url#1{{\tt #1}}\fi
\expandafter\ifx\csname urlprefix\endcsname\relax\def\urlprefix{URL }\fi
\ifx\endbibitem\undefined \let\endbibitem\relax\fi

\bibitem[{Allen(2017)}]{Allen2017}
Allen, L. J.~S. (2017).
\newblock \enquote{A primer on stochastic epidemic models: Formulation,
  numerical simulation, and analysis.}
\newblock {\em Infectious Disease Modelling\/}, 2: 128--142.
\newline\urlprefix\url{http://www.ncbi.nlm.nih.gov/pubmed/29928733
  http://www.pubmedcentral.nih.gov/articlerender.fcgi?artid=PMC6002090}
\endbibitem

\bibitem[{Ashall et~al.(2009)Ashall, Horton, Nelson, Paszek, Harper, Sillitoe,
  Ryan, Spiller, Unitt, Broomhead, Kell, Rand, S{\'e}e, and White}]{ashall}
Ashall, L., Horton, C.~A., Nelson, D.~E., Paszek, P., Harper, C.~V., Sillitoe,
  K., Ryan, S., Spiller, D.~G., Unitt, J.~F., Broomhead, D.~S., Kell, D.~B.,
  Rand, D.~A., S{\'e}e, V., and White, M. R.~H. (2009).
\newblock \enquote{Pulsatile Stimulation Determines Timing and Specificity of
  {NF}-{$\kappa$}{B}-Dependent Transcription.}
\newblock {\em Science\/}, 324(5924): 242--246.
\endbibitem

\bibitem[{Baggenstoss(2018)}]{baggenstoss}
Baggenstoss, P.~M. (2018).
\newblock \enquote{Acoustic Event Classification Using Multi-Resolution HMM.}
\newblock In {\em 2018 26th European Signal Processing Conference (EUSIPCO)\/},
  972--976.
\endbibitem

\bibitem[{Beaumont(2003)}]{beaumontabc}
Beaumont, M.~A. (2003).
\newblock \enquote{{Estimation of Population Growth or Decline in Genetically
  Monitored Populations}.}
\newblock {\em Genetics\/}, 164(3): 1139--1160.
\endbibitem

\bibitem[{Boettiger(2018)}]{Boettiger2018}
Boettiger, C. (2018).
\newblock \enquote{From noise to knowledge: how randomness generates novel
  phenomena and reveals information.}
\newblock {\em Ecology Letters\/}, 21: 1255--1267.
\newline\urlprefix\url{http://doi.wiley.com/10.1111/ele.13085}
\endbibitem

\bibitem[{Browning et~al.(2020)Browning, Warne, Burrage, Baker, and
  Simpson}]{browning2020}
Browning, A.~P., Warne, D.~J., Burrage, K., Baker, R.~E., and Simpson, M.~J.
  (2020).
\newblock \enquote{Identifiability analysis for stochastic differential
  equation models in systems biology.}
\newblock {\em Journal of The Royal Society Interface\/}, 17(173): 20200652.
\newline\urlprefix\url{https://royalsocietypublishing.org/doi/abs/10.1098/rsif.2020.0652}
\endbibitem

\bibitem[{Burton et~al.(2021)Burton, Manning, Rattray, Papalopulu, and
  Kursawe}]{burton2021}
Burton, J., Manning, C.~S., Rattray, M., Papalopulu, N., and Kursawe, J.
  (2021).
\newblock \enquote{Inferring kinetic parameters of oscillatory gene regulation
  from single cell time-series data.}
\newblock {\em Journal of The Royal Society Interface\/}, 18(182): 20210393.
\endbibitem

\bibitem[{Chib and Greenberg(1995)}]{chib}
Chib, S. and Greenberg, E. (1995).
\newblock \enquote{Understanding the Metropolis-Hastings Algorithm.}
\newblock {\em American Statistician\/}, 49: 327--335.
\endbibitem

\bibitem[{Cover and Thomas(2006)}]{CoverThomas}
Cover, T.~M. and Thomas, J.~A. (2006).
\newblock {\em Elements of Information Theory (Wiley Series in
  Telecommunications and Signal Processing)\/}.
\newblock USA: Wiley-Interscience.
\endbibitem

\bibitem[{DeFelice et~al.(2019)DeFelice, Clark, Hughey, Maayan, Kudo, Gutschow,
  Covert, and Regot}]{DeFelice2019}
DeFelice, M.~M., Clark, H.~R., Hughey, J.~J., Maayan, I., Kudo, T., Gutschow,
  M.~V., Covert, M.~W., and Regot, S. (2019).
\newblock \enquote{NF-$\kappa$B signaling dynamics is controlled by a
  dose-sensing autoregulatory loop.}
\newblock {\em Science Signaling\/}, 12.
\endbibitem

\bibitem[{Earl and Deem(2005)}]{ptMCMCclassic}
Earl, D.~J. and Deem, M.~W. (2005).
\newblock \enquote{Parallel tempering: Theory{,} applications{,} and new
  perspectives.}
\newblock {\em Phys. Chem. Chem. Phys.\/}, 7: 3910--3916.
\newline\urlprefix\url{http://dx.doi.org/10.1039/B509983H}
\endbibitem

\bibitem[{Fearnhead et~al.(2014)Fearnhead, Giagos, and Sherlock}]{fearnhead14}
Fearnhead, P., Giagos, V., and Sherlock, C. (2014).
\newblock \enquote{Inference for reaction networks using the linear noise
  approximation.}
\newblock {\em Biometrics\/}, 70(2): 457--466.
\endbibitem

\bibitem[{Finkenst{\"a}dt et~al.(2013)Finkenst{\"a}dt, Woodcock, Komorowski,
  Harper, Davis, White, and Rand}]{finkenstadt13}
Finkenst{\"a}dt, B., Woodcock, D.~J., Komorowski, M., Harper, C.~V., Davis, J.
  R.~E., White, M. R.~H., and Rand, D.~A. (2013).
\newblock \enquote{Quantifying intrinsic and extrinsic noise in gene
  transcription using the linear noise approximation: an application to single
  cell data.}
\newblock {\em The Annals of Applied Statistics\/}, 7(4): 1960--1982.
\endbibitem

\bibitem[{Forger(2017)}]{Forgerbook}
Forger, D.~B. (2017).
\newblock {\em Biological clocks, rhythms, and oscillations: The theory of
  biological timekeeping\/}.
\newblock Cambridge (MA): MIT Press. MIT Press, Cambridge (MA).
\endbibitem

\bibitem[{Froda and Nkurunziza(2007)}]{froda07}
Froda, S. and Nkurunziza, S. (2007).
\newblock \enquote{Prediction of predator--prey populations modelled by
  perturbed ODEs.}
\newblock {\em Journal of Mathematical Biology\/}, 54(3): 407--451.
\newline\urlprefix\url{https://doi.org/10.1007/s00285-006-0051-9}
\endbibitem

\bibitem[{Gabriel et~al.(2021)Gabriel, del Olmo, Zehtabian, J{\"a}ger, Reischl,
  van Dijk, Ulbricht, Rakhymzhan, Korte, Koller, Grudziecki, Maier, Herrmann,
  Niesner, Zemojtel, Ewers, Granada, Herzel, and Kramer}]{Gabriel2021}
Gabriel, C.~H., del Olmo, M., Zehtabian, A., J{\"a}ger, M., Reischl, S., van
  Dijk, H., Ulbricht, C., Rakhymzhan, A., Korte, T., Koller, B., Grudziecki,
  A., Maier, B., Herrmann, A., Niesner, R., Zemojtel, T., Ewers, H., Granada,
  A.~E., Herzel, H., and Kramer, A. (2021).
\newblock \enquote{Live-cell imaging of circadian clock protein dynamics in
  CRISPR-generated knock-in cells.}
\newblock {\em Nature Communications\/}, 12: 3796.
\endbibitem

\bibitem[{Gard and Kannan(1976)}]{gard_kannan_1976}
Gard, T.~C. and Kannan, D. (1976).
\newblock \enquote{On a stochastic differential equation modeling of
  prey-predator evolution.}
\newblock {\em Journal of Applied Probability\/}, 13(3): 429--443.
\endbibitem

\bibitem[{Gillespie(1977)}]{gillespe}
Gillespie, D.~T. (1977).
\newblock \enquote{Exact stochastic simulation of coupled chemical reactions.}
\newblock {\em The Journal of Physical Chemistry\/}, 81(25): 2340--2361.
\endbibitem

\bibitem[{Gillespie(1992)}]{Gillespie1992}
--- (1992).
\newblock \enquote{A rigorous derivation of the chemical master equation.}
\newblock {\em Physica A: Statistical Mechanics and its Applications\/}, 188:
  404--425.
\newline\urlprefix\url{http://www.sciencedirect.com/science/article/pii/037843719290283V
  https://www.sciencedirect.com/science/article/pii/037843719290283V}
\endbibitem

\bibitem[{Gillespie(2000)}]{Gillespie2000a}
--- (2000).
\newblock \enquote{The chemical Langevin equation.}
\newblock {\em The Journal of Chemical Physics\/}, 113: 297--306.
\newline\urlprefix\url{http://scitation.aip.org/content/aip/journal/jcp/113/1/10.1063/1.481811}
\endbibitem

\bibitem[{Gillespie and Petzold(2003)}]{Gillespie2003}
Gillespie, D.~T. and Petzold, L.~R. (2003).
\newblock \enquote{Improved leap-size selection for accelerated stochastic
  simulation.}
\newblock {\em The Journal of Chemical Physics\/}, 119.
\endbibitem

\bibitem[{Girolami and Calderhead(2011)}]{girolami11}
Girolami, M. and Calderhead, B. (2011).
\newblock \enquote{Riemann manifold Langevin and Hamiltonian Monte Carlo
  methods.}
\newblock {\em Journal of the Royal Statistical Society: Series B (Statistical
  Methodology)\/}, 73(2): 123--214.
\endbibitem

\bibitem[{Goldental et~al.(2017)Goldental, Uzan, Sardi, and
  Kanter}]{Goldental2017}
Goldental, A., Uzan, H., Sardi, S., and Kanter, I. (2017).
\newblock \enquote{Oscillations in networks of networks stem from adaptive
  nodes with memory.}
\newblock {\em Scientific Reports\/}, 7: 2700.
\endbibitem

\bibitem[{Golightly and Wilkinson(2011)}]{Golightly2011}
Golightly, A. and Wilkinson, D.~J. (2011).
\newblock \enquote{Bayesian parameter inference for stochastic biochemical
  network models using particle Markov chain Monte Carlo.}
\newblock {\em Interface Focus\/}, 1: 807--820.
\endbibitem

\bibitem[{Gonze et~al.(2003)Gonze, Halloy, Leloup, and Goldbeter}]{drosmodel}
Gonze, D., Halloy, J., Leloup, J.-C., and Goldbeter, A. (2003).
\newblock \enquote{Stochastic models for circadian rhythms: effect of molecular
  noise on periodic and chaotic behaviour.}
\newblock {\em Comptes Rendus Biologies\/}, 326(2): 189 -- 203.
\endbibitem

\bibitem[{Gonze and Ruoff(2021)}]{Gonze2021}
Gonze, D. and Ruoff, P. (2021).
\newblock \enquote{The Goodwin Oscillator and its Legacy.}
\newblock {\em Acta Biotheoretica\/}, 69: 857--874.
\newline\urlprefix\url{http://link.springer.com/10.1007/s10441-020-09379-8}
\endbibitem

\bibitem[{Gramacy et~al.(2010)Gramacy, Samworth, and King}]{gramacy10}
Gramacy, R., Samworth, R., and King, R. (2010).
\newblock \enquote{Importance tempering.}
\newblock {\em Statistics and Computing\/}, 20(1): 1--7.
\newline\urlprefix\url{https://doi.org/10.1007/s11222-008-9108-5}
\endbibitem

\bibitem[{Greer et~al.(2020)Greer, Saha, Gogliettino, Yu, and
  Zollo-Venecek}]{Greer2020}
Greer, M., Saha, R., Gogliettino, A., Yu, C., and Zollo-Venecek, K. (2020).
\newblock \enquote{Emergence of oscillations in a simple epidemic model with
  demographic data.}
\newblock {\em Royal Society Open Science\/}, 7: 191187.
\endbibitem

\bibitem[{Grewal and Andrews(2014)}]{KalmanFilter_book}
Grewal, M.~S. and Andrews, A.~P. (eds.) (2014).
\newblock {\em Kalman Filtering: Theory and Practice with MATLAB (4th ed.)\/}.
\newblock Wiley-IEEE Press.
\endbibitem

\bibitem[{Grima et~al.(2011)Grima, Thomas, and Straube}]{Grima2011}
Grima, R., Thomas, P., and Straube, A.~V. (2011).
\newblock \enquote{How accurate are the nonlinear chemical Fokker-Planck and
  chemical Langevin equations?}
\newblock {\em The Journal of Chemical Physics\/}, 135: 084103.
\endbibitem

\bibitem[{Gupta et~al.(2018)Gupta, Hainsworth, Hogg, Lee, and Faeder}]{gupta18}
Gupta, S., Hainsworth, L., Hogg, J., Lee, R., and Faeder, J. (2018).
\newblock \enquote{Evaluation of Parallel Tempering to Accelerate Bayesian
  Parameter Estimation in Systems Biology.}
\newblock In {\em 2018 26th Euromicro International Conference on Parallel,
  Distributed and Network-based Processing (PDP)\/}, 690--697.
\endbibitem

\bibitem[{Gutenkunst et~al.(2007)Gutenkunst, Waterfall, Casey, Brown, Myers,
  and Sethna}]{Gutenkunst:2007}
Gutenkunst, R.~N., Waterfall, J.~J., Casey, F.~P., Brown, K.~S., Myers, C.~R.,
  and Sethna, J.~P. (2007).
\newblock \enquote{Universally Sloppy Parameter Sensitivities in Systems
  Biology Models.}
\newblock {\em PLoS Comput Biol\/}, 3: e189.
\newline\urlprefix\url{https://dx.plos.org/10.1371/journal.pcbi.0030189}
\endbibitem

\bibitem[{Hansmann(1997)}]{hansmann}
Hansmann, U.~H. (1997).
\newblock \enquote{Parallel tempering algorithm for conformational studies of
  biological molecules.}
\newblock {\em Chemical Physics Letters\/}, 281(1): 140 -- 150.
\newline\urlprefix\url{http://www.sciencedirect.com/science/article/pii/S0009261497011986}
\endbibitem

\bibitem[{Harper et~al.(2018)Harper, Woodcock, Lam, Garcia-Albornoz, Adamson,
  Ashall, Rowe, Downton, Schmidt, West, Spiller, Rand, and White}]{Harper2018a}
Harper, C.~V., Woodcock, D.~J., Lam, C., Garcia-Albornoz, M., Adamson, A.,
  Ashall, L., Rowe, W., Downton, P., Schmidt, L., West, S., Spiller, D.~G.,
  Rand, D.~A., and White, M. R.~H. (2018).
\newblock \enquote{Temperature regulates NF-$\kappa$B dynamics and function
  through timing of A20 transcription.}
\newblock {\em Proceedings of the National Academy of Sciences\/}.
\newline\urlprefix\url{https://www.pnas.org/content/pnas/115/22/E5243.full.pdf}
\endbibitem

\bibitem[{Ito and Uchida(2010)}]{Ito2010}
Ito, Y. and Uchida, K. (2010).
\newblock \enquote{Formulas for intrinsic noise evaluation in oscillatory
  genetic networks.}
\newblock {\em Journal of Theoretical Biology\/}, 267: 223--234.
\newline\urlprefix\url{http://www.sciencedirect.com/science/article/pii/S0022519310004480}
\endbibitem

\bibitem[{Komorowski et~al.(2009)Komorowski, Finkenst{\"a}dt, Harper, and
  Rand}]{Komorowski2009}
Komorowski, M., Finkenst{\"a}dt, B., Harper, C.~V., and Rand, D.~A. (2009).
\newblock \enquote{Bayesian inference of biochemical kinetic parameters using
  the linear noise approximation.}
\newblock {\em BMC Bioinformatics\/}, 10(1): 343.
\endbibitem

\bibitem[{Kurtz(1970)}]{kurtz70}
Kurtz, T.~G. (1970).
\newblock \enquote{Solutions of ordinary differential equations as limits of
  pure jump markov processes.}
\newblock {\em Journal of Applied Probability\/}, 7(1): 49--58.
\endbibitem

\bibitem[{Kurtz(1971)}]{kurtz71}
--- (1971).
\newblock \enquote{Limit theorems for sequences of jump Markov processes
  approximating ordinary differential processes.}
\newblock {\em Journal of Applied Probability\/}, 8(2): 344--356.
\endbibitem

\bibitem[{Lane et~al.(2017)Lane, Valen, DeFelice, Macklin, Kudo, Jaimovich,
  Carr, Meyer, Pe'er, Boutet, and Covert}]{Lane2017}
Lane, K., Valen, D.~V., DeFelice, M.~M., Macklin, D.~N., Kudo, T., Jaimovich,
  A., Carr, A., Meyer, T., Pe'er, D., Boutet, S.~C., and Covert, M.~W. (2017).
\newblock \enquote{Measuring Signaling and RNA-Seq in the Same Cell Links Gene
  Expression to Dynamic Patterns of NF-$\kappa$B Activation.}
\newblock {\em Cell Systems\/}, 4: 458--469.e5.
\endbibitem

\bibitem[{Lei(2021)}]{sysbio}
Lei, J. (2021).
\newblock {\em Systems biology : modeling, analysis, and simulation\/}.
\newblock Lecture notes on mathematical modelling in the life sciences.
  Springer Cham.
\endbibitem

\bibitem[{Marinopoulou et~al.(2021)Marinopoulou, Biga, Sabherwal, Miller,
  Desai, Adamson, and Papalopulu}]{Marinopoulou2021}
Marinopoulou, E., Biga, V., Sabherwal, N., Miller, A., Desai, J., Adamson,
  A.~D., and Papalopulu, N. (2021).
\newblock \enquote{HES1 protein oscillations are necessary for neural stem
  cells to exit from quiescence.}
\newblock {\em iScience\/}, 24: 103198.
\endbibitem

\bibitem[{McCrea et~al.(2023)McCrea, King, Graham, and Börger}]{mccrea2023}
McCrea, R., King, R., Graham, L., and Börger, L. (2023).
\newblock \enquote{Realising the promise of large data and complex models.}
\newblock {\em Methods in Ecology and Evolution\/}, 14(1): 4--11.
\newline\urlprefix\url{https://besjournals.onlinelibrary.wiley.com/doi/abs/10.1111/2041-210X.14050}
\endbibitem

\bibitem[{Minas and Rand(2017)}]{minasrand17}
Minas, G. and Rand, D.~A. (2017).
\newblock \enquote{Long-time analytic approximation of large stochastic
  oscillators: Simulation, analysis and inference.}
\newblock {\em PLOS Computational Biology\/}, 13(7): 1--23.
\endbibitem

\bibitem[{Minas and Rand(2019)}]{Minas2019}
--- (2019).
\newblock \enquote{Parameter sensitivity analysis for biochemical reaction
  networks.}
\newblock {\em Mathematical Biosciences and Engineering\/}, 16: 3965--3987.
\newline\urlprefix\url{http://www.aimspress.com/article/10.3934/mbe.2019196}
\endbibitem

\bibitem[{Rand(2008)}]{Rand2008}
Rand, D.~A. (2008).
\newblock \enquote{Mapping global sensitivity of cellular network dynamics:
  sensitivity heat maps and a global summation law.}
\newblock {\em Journal of The Royal Society Interface\/}, 5: S59--S69.
\newline\urlprefix\url{http://rsif.royalsocietypublishing.org/content/5/Suppl_1/S59.abstract}
\endbibitem

\bibitem[{Schnoerr et~al.(2017)Schnoerr, Sanguinetti, and Grima}]{Schnoerr}
Schnoerr, D., Sanguinetti, G., and Grima, R. (2017).
\newblock \enquote{Approximation and inference methods for stochastic
  biochemical kinetics{\textemdash}a tutorial review.}
\newblock {\em Journal of Physics A: Mathematical and Theoretical\/}, 50(9):
  093001.
\endbibitem

\bibitem[{Sisson et~al.(2018)Sisson, Fan, and Beaumont}]{abc_book}
Sisson, S.~A., Fan, Y., and Beaumont, M.~A. (eds.) (2018).
\newblock {\em Handbook of Approximate Bayesian Computation\/}.
\newblock Chapman and Hall/CRC.
\endbibitem

\bibitem[{Spitschan et~al.(2024)Spitschan, Zauner, {Nilsson Tengelin},
  Bouroussis, Caspar, and Eloi}]{spitschan2024}
Spitschan, M., Zauner, J., {Nilsson Tengelin}, M., Bouroussis, C.~A., Caspar,
  P., and Eloi, F. (2024).
\newblock \enquote{Illuminating the future of wearable light metrology:
  Overview of the MeLiDos Project.}
\newblock {\em Measurement\/}, 235: 114909.
\newline\urlprefix\url{https://www.sciencedirect.com/science/article/pii/S0263224124007942}
\endbibitem

\bibitem[{Tay et~al.(2010{\natexlab{a}})Tay, Hughey, Lee, Lipniacki, Quake, and
  Covert}]{Tay2010}
Tay, S., Hughey, J.~J., Lee, T.~K., Lipniacki, T., Quake, S.~R., and Covert,
  M.~W. (2010{\natexlab{a}}).
\newblock \enquote{Single-cell NF-B dynamics reveal digital activation and
  analogue information processing.}
\newblock {\em Nature\/}, 466: 267--271.
\endbibitem

\bibitem[{Tay et~al.(2010{\natexlab{b}})Tay, Hughey, Lee, Lipniacki, Quake, and
  Covert}]{Tay}
--- (2010{\natexlab{b}}).
\newblock \enquote{Single-cell NF-$\kappa$B dynamics reveal digital activation
  and analogue information processing.}
\newblock {\em Nature\/}, 466(7303): 267--271.
\newline\urlprefix\url{https://doi.org/10.1038/nature09145}
\endbibitem

\bibitem[{Van~Kampen(2007)}]{vankampen92}
Van~Kampen, N.~G. (2007).
\newblock {\em Stochastic Processes in Physics and Chemistry\/}.
\newblock North-Holland Personal Library. Amsterdam: Elsevier, third edition
  edition.
\endbibitem

\bibitem[{Vehtari et~al.(2021)Vehtari, Gelman, Simpson, Carpenter, and
  B{\"u}rkner}]{vehtari21}
Vehtari, A., Gelman, A., Simpson, D., Carpenter, B., and B{\"u}rkner, P.-C.
  (2021).
\newblock \enquote{{Rank-Normalization, Folding, and Localization: An Improved
  $\widehat{R}$ for Assessing Convergence of MCMC (with Discussion)}.}
\newblock {\em Bayesian Analysis\/}, 16(2): 667 -- 718.
\newline\urlprefix\url{https://doi.org/10.1214/20-BA1221}
\endbibitem

\bibitem[{Wallace et~al.(2012)Wallace, Gillespie, Sanft, Petzold, Gillespie,
  and Sanft}]{Wallace2012}
Wallace, E. W.~J., Gillespie, D.~T., Sanft, K.~R., Petzold, L.~R., Gillespie,
  D.~T., and Sanft, K.~R. (2012).
\newblock \enquote{Linear noise approximation is valid over limited times for
  any chemical system that is sufficiently large.}
\newblock {\em IET SYSTEMS BIOLOGY\/}, 6: 102--115.
\newline\urlprefix\url{https://digital-library.theiet.org/content/journals/10.1049/iet-syb.2011.0038}
\endbibitem

\bibitem[{Weitz et~al.(2020)Weitz, Park, Eksin, and Dushoff}]{Weitz2020}
Weitz, J.~S., Park, S.~W., Eksin, C., and Dushoff, J. (2020).
\newblock \enquote{Awareness-driven behavior changes can shift the shape of
  epidemics away from peaks and toward plateaus, shoulders, and oscillations.}
\newblock {\em Proceedings of the National Academy of Sciences\/}, 117:
  32764--32771.
\endbibitem

\bibitem[{Wilkinson(2011)}]{wilkinsonsmsb}
Wilkinson, D. (2011).
\newblock {\em Stochastic Modelling for Systems Biology, Second Edition\/}.
\newblock Chapman \& Hall/CRC Mathematical and Computational Biology. Taylor \&
  Francis.
\endbibitem

\bibitem[{Zhang et~al.(2017)Zhang, Lenardo, and Baltimore}]{Zhang2017d}
Zhang, Q., Lenardo, M.~J., and Baltimore, D. (2017).
\newblock \enquote{30 Years of NF-$\kappa$B: A Blossoming of Relevance to Human
  Pathobiology.}
\newblock {\em Cell\/}, 168: 37--57.
\newline\urlprefix\url{https://www.sciencedirect.com/science/article/pii/S0092867416317263}
\endbibitem

\end{thebibliography}


\begin{thebibliography}{0}
\newcommand{\enquote}[1]{``#1''}
\expandafter\ifx\csname natexlab\endcsname\relax\def\natexlab#1{#1}\fi
\expandafter\ifx\csname url\endcsname\relax
  \def\url#1{{\tt #1}}\fi
\expandafter\ifx\csname urlprefix\endcsname\relax\def\urlprefix{URL }\fi
\ifx\endbibitem\undefined \let\endbibitem\relax\fi

\end{thebibliography}








\end{document}


\begin{frontmatter}
\title{Supplementary information for ``Bayesian inference for stochastic oscillatory systems using the phase-corrected Linear Noise Approximation''\thanksref{T1}}
\runtitle{Bayesian inference for oscillatory systems}

\begin{aug}
\author{\fnms{Ben} \snm{Swallow}\thanksref{addr1,t1,t2,m1}\ead[label=e1]{bts3@st-andrews.ac.uk}},
\author{\fnms{David A.} \snm{Rand}\thanksref{addr2,t1,m2}%
\ead[label=e2]{d.a.rand@warwick.ac.uk}}
\and
\author{\fnms{Giorgos} \snm{Minas}\thanksref{addr3,t3,m1,m2}\ead[label=e3]{gm256@st-andrews.ac.uk}}

\runauthor{B. Swallow et al.}

\address[addr1]{School of Mathematics and Statistics, University of St Andrews, St Andrews, UK,   
    \printead{e1}, 
}

\address[addr2]{Mathematics Institute, University of Warwick, Coventry, UK
    \printead{e2}
}

\address[addr3]{School of Mathematics and Statistics, University of St Andrews, St Andrews, UK,
\printead{e3}}


\end{aug}

\begin{abstract}
    This document contains supplementary information for the publication ``Bayesian inference for stochastic oscillatory systems using the phase-corrected Linear Noise Approximation''
\end{abstract}



\end{frontmatter}

\section{Details of the Kalman Filter}

We next go through the details of the pcLNA Kalman Filter steps. Since the same steps are used for $i=0,1,\dots,n$, we treat $i=0$ as an example and provide derivations in detail, then noting how to generalise for $i=1,2,\dots,n$. 

\paragraph{KF2 and KF3d} 

We have that $(\bx_0|\by_{-1})  \sim N(\bmu_0, \bSigma_0)$ and hence the observation equation, for $i=0,1,\dots,N$,
$$\by_i = \by(t_i) = \bB \bx_i + \bepsilon_i,\quad \bepsilon_i \sim N(0,\bSigma_e),\quad \bepsilon_i \text{ independent to } \bx_i $$ 
implies the distribution of $( \by_0| \by_{-1})\sim N( \bmu^{(\by)}_0 ,\bSigma^{(\by)}_0)$, where $\bmu^{(\by)}_0 = \bB \bmu_0 $, $\bSigma^{(\by)}_0 = \bB\bSigma_0\bB^T + \bSigma_e$

Similarly for KF3d, we have that $(\bx_i|\by_{i-1})  \sim N(\bmu_i, \bSigma_i)$ and using the observation equation we get $( \by_i| \by_{i-1})\sim N( \bmu^{(\by)}_i ,\bSigma^{(\by)}_i)$, where $\bmu^{(\by)}_i = \bB \bmu_i $, $\bSigma^{(\by)}_i = \bB\bSigma_i\bB^T + \bSigma_e$

\paragraph{KF3a}
Using the covariance, $i=0,1,\dots,N$
$$\Cov(\bx_{i},\by_{i} ) = \Cov(\bx_{i},\bB\bx_i + \bepsilon_i )=\Cov(\bx_{i},\bB\bx_i)=\bB \;\Var(\bx_{i} )$$ and hence, using the findings in KF2, we get the joint distribution 
\begin{equation}\label{eq:jointPrior0} 
\left( \left. \left( \begin{matrix} \bx_{0}\\ \by_{0} \end{matrix}\right) \right|  \by_{-1}  \right) \sim N\left( \left( \begin{matrix} \bmu_{0}\\ \bB \bmu_0 \end{matrix}\right), \left( \begin{matrix} \bSigma_{0} & \bSigma_0\bB^T \\ \bB \bSigma_0 & \bSigma_0^{(y)} \end{matrix}\right) \right)  .
\end{equation}
Then, $(\bx_0|\bY_{(0})$ is derived as the conditional distribution of the first component, $\bx_0$, above given the second component, $\by_0$ (and $\by_{-1}$), of the MVN distribution in \eqref{eq:jointPrior0}. That is, 
$(\bx_{0}| \bY_{(0)}) \sim N(\bmu^*_{0},\bSigma^*_{0})$ where
\begin{eqnarray*}
 \bmu^*_{0} &=& \bmu_{0} + \bSigma_{0} \bB^T (\bSigma^{(\by)}_{0})^{-1} \hat{\bepsilon}_{0}, \quad \hat{\bepsilon}_{0} = \by_{0} - \bmu^\by_{0}\\
 \bSigma^*_{0}&=& \bSigma_{0} - \bSigma_{0}\bB^T (\bSigma^{(\by)}_{0})^{-1} \bB\bSigma_{0}.
\end{eqnarray*} \label{KF:step3a}

Similarly, for $i=2,\dots,N$, using KF3d we have that
\begin{equation}\label{eq:jointPriori} 
\left( \left. \left( \begin{matrix} \bx_{i-1}\\ \by_{i-1} \end{matrix}\right) \right|  \bY_{(i-2)}  \right) \sim N\left( \left( \begin{matrix} \bmu_{i-1}\\ \bB \bmu_{i-1} \end{matrix}\right), \left( \begin{matrix} \bSigma_{i-1} & \bSigma_{i-1}\bB^T \\ \bB \bSigma_{i-1} & \bSigma_{i-1}^{(y)} \end{matrix}\right) \right)  .
\end{equation}
and then 
$(\bx_{i-1}|\bY_{(i-1)})$ is derived as the conditional distribution of the first component, $\bx_{i-1}$, above given the second component, $\by_{i-1}$ (and $\bY_{(i-2)}$), of the MVN distribution in \eqref{eq:jointPriori}. That is, $(\bx_{i-1}| \bY_{(i-1)}) \sim N(\bmu^*_{i-1},\bSigma^*_{i-1})$ where
\begin{eqnarray*}
 \bmu^*_{i-1} &=& \bmu_{i-1} + \bSigma_{i-1} \bB^T (\bSigma^{(\by)}_{i-1})^{-1} \hat{\bepsilon}_{i-1}, \quad \hat{\bepsilon}_{i-1} = \by_{i-1} - \bmu^\by_{i-1}\\
 \bSigma^*_{i-1}&=& \bSigma_{i-1} - \bSigma_{i-1}\bB^T (\bSigma^{(\by)}_{i-1})^{-1} \bB\bSigma_{i-1}.
\end{eqnarray*} \label{KF:step3a}

\paragraph{KF3b - phase correction}


The phase $s_{i-1}$, $i=1,\dots,N$ that gives the closest point $\bphi(s_{i-1})$ of the ODE solution to the scaled mean $\bmu^*_{i-1}/\Omega$ satisfies $$\bF(\bphi(s_{i-1}))^T(\bmu^*_{i-1}/\Omega - \bphi(s_{i-1}))=0.$$ 
That is, the vector $\bmu^*_{i-1}/\Omega - \bphi(s_{i-1})$ is orthogonal to the tangent vector $\bF(\bphi(s_{i-1}))$.  
Therefore, our target is to find a root of the score function 
\[ \text{score}(s) = \bF(\bphi(s))^T(\bmu^*_{i-1}/\Omega - \bphi(s)). \]
The first derivative of the score function is
\begin{align}
\text{score}'(s) &= \bF'(\bphi(s))^T(\bmu^*_{i-1}/\Omega - \bphi(s)) -\bF(\bphi(s))^T \bphi'(s) \\
&= \bJ(\bphi(s))\bF(\bphi(s))(\bmu^*_{i-1}/\Omega - \bphi(s)) - \bF(\bphi(s))^T\bF(\bphi(s)),
\end{align}
where in the last equation we used that $\bF'(\bphi(s))=\bJ(\bphi(s))\bF(\bphi(s))$.

Starting with $s_{i-1}^{(0)}$, we compute 
$$s^{(k+1)}_i = s^{(k)}_i - \text{score}\left(s^{(k)}_i \right)/\text{score}'\left(s^{(k)}_i \right),$$
for $k=1,2,\dots,k^*$, where $k^*$ the smallest $k$ such that $\left|\text{score}(s^{(k^*)}_i)\right|<c$, for some small constant $c>0$. We then set $s_i=s_i^{k^*}$ and $\bphi_i = \bphi(s_i)$. This is the standard Newton-Raphson method, but other methods for minimising the score function can be used. 

The phase-corrected noise $\bkappa_{0}$ satisfies the pcLNA ansantz equation 
\[ \bx_0 = \Omega \bphi_0 + \sqrt{\Omega} \bkappa_0 \]
and its tangential coordinate is restricted to be $0$. 
To derive its distribution, we apply a change of coordinates to have the first coordinate being parallel to the tangential direction, $\bF(\bphi_0)$, at $\bphi_0$, and derive the conditional distribution on the transversal section given that the tangential direction is set to $0$.
More specifically, let $\tilde{\bkappa}_0 = (\bx_0/\Omega - \bphi_0)/\sqrt{\Omega}$. 
Then, $(\tilde{\bkappa}_0| \bY_{0} ) \sim N(\tilde{\bom}_0^*,\tilde{\bS}_0^*) $, where
$\tilde{\bom}_0^* =(\bmu_0^*/\Omega - \bphi_0)/\sqrt{\Omega}$, $\tilde{\bS}_0^* = \bSigma^*_0/\Omega.$ Let $\bE = [\be_1\, \bE_2]$ where $\be_1 :=  \bF(\bphi_{i-1})/\|\bF(\bphi_{i-1})\|$, and $\bE_2:=[\be_2 \cdots \be_n]$ with $\be_{j}^T \be^{}_{j}=1$, and $\be_{j}^T \be^{}_{j'}=0$, for $j\neq j'$,  and $j,j'\in \{1,2,\dots,N\}$. Then, because $\bphi_0$ is chosen to minimise the Euclidean distance $d(\bmu^*_0/\Omega,\bphi(s))$, $\tilde{\bom}_0^*$ is orthogonal to the tangent vector $\bF(\bphi_0)$, and $\be_1^T\tilde{\bom}_0^*=0 $. 
We derive that $(\bE^T \tilde{\bkappa}_0|\bY_{(0)} ) \sim N(\bE^T\tilde{\bom}_0^*,\bE^T\tilde{\bS}_0^*\bE), $
with $$\bE^T\tilde{\bom}_0^* = \left( \begin{matrix}
     0 \\ \bE_2^T\tilde{\bom}_0^*\end{matrix} \right), \qquad  \bE^T\tilde{\bS}_0^*\bE = \left(\begin{matrix}
    \be_1^T\tilde{\bS}_0^*\be_1 & \be_1^T\tilde{\bS}_0^*\bE_2 \\ \bE_2^T \tilde{\bS}_0^*\be_1 & \bE_2^T \tilde{\bS}_0^*\bE_2
\end{matrix} \right)$$
and the (phase) conditional distribution 
$$(\bE_2^T \tilde{\bkappa}_0 | \be_1^T \tilde{\bkappa}_0 =0, \bY_{(0)}) \sim N\left(\bE_2^T\tilde{\bom}_0^*, \bE_2^T\tilde{\bS}_0^*\bE_2 - \frac{(\be_1^T \tilde{\bS}_0^* \bE_2 )^T(\be_1^T \tilde{\bS}_0^* \bE_2 )}{ (\be_1^T \tilde{\bS}_0^* \be_1 )} \right).$$
Then to move the distribution back to the original coordinates, we get
$$ \left( \left. \bkappa_0 := \bE\left(\begin{matrix}
    0 \\ \bE_2^T\tilde{\bkappa}_0^* 
\end{matrix}  \right) \right| \bY_{(0)} \right) \sim N\left( \bom_0^*,\bS_0^* \right) $$
where $$\bom_0^* =\bE\left(\begin{matrix}
    0 \\ \bE_2^T\tilde{\bom}_0^* 
\end{matrix}  \right)= \bE\left(\begin{matrix}
    \be_1^T\bom_0^*  \\ \bE_2^T\tilde{\bom}_0^* 
\end{matrix}  \right) = \bE \bE^T\tilde{\bom}_0^*=\tilde{\bom}_0^* =(\bmu_0^*/\Omega - \bphi_0)/\sqrt{\Omega} $$
and
\begin{align*}
    \bS_0^* &= \bE\left(\begin{matrix}
    0 & {\bm 0}^T \\ {\bm 0} & \bE_2^T\tilde{\bS}_0^*\bE_2 - \frac{(\be_1^T \tilde{\bS}_0^* \bE_2 )^T(\be_1^T \tilde{\bS}_0^* \bE_2 )}{ \be_1^T \tilde{\bS}_0^* \be_1 } \end{matrix} \right) \bE^T \\ 
    &= \bE_2 \left(\bE_2^T\tilde{\bS}_0^*\bE_2 - \frac{(\be_1^T \tilde{\bS}_0^* \bE_2 )^T(\be_1^T \tilde{\bS}_0^* \bE_2 )}{ \be_1^T \tilde{\bS}_0^* \be_1}\right)\bE_2^T \\
    &= \Omega \bE_2 \left(\bE_2^T\tilde{\bSigma}_0^*\bE_2 - \frac{(\be_1^T \tilde{\bSigma}_0^* \bE_2 )^T(\be_1^T \tilde{\bSigma}_0^* \bE_2 )}{ \be_1^T \tilde{\bSigma}_0^* \be_1}\right)\bE_2^T.
\end{align*}

To derive the distribution of $(\bkappa_{i-1} | \bY_{(i-1)}) \sim N(\bom_{i-1}^*,\bS_{i-1}^*) $, for $i=2,\dots,N$, we perform the same steps only with $\bphi_0$ replaced with $\bphi_{i-1}$ and thus a different matrix $\bE=[\be_1 \, \bE_2]$ is used, now with $\be_1$ being parallel to $\bF(\bphi_{i-1})$. 

\paragraph{KF3c}
For $i=1,2,\dots,N$, using the transition equation (2.8)  
$$\bphi_{i} = \bphi(s_{i-1}\! +\!\Delta t_{i}),\qquad \Delta t_{i} = t_{i}-t_{i-1}$$
and (2.9) 
that gives
$$ \bxi_{i} = \bC(s_{i-1},s_{i-1}\! +\!\Delta t_{i})\, \bkappa_{i-1}+ \boeta_{i} ,  \qquad \boeta_{i} \sim N(0,\bV(s_{i-1},s_{i-1}\! +\!\Delta t_{i})).$$
we get the state equation as in (2.7) 
$$\bx_{i} =   \bx(t_{i}) =\Omega \bphi_{i}\! +\! \sqrt{\Omega}\, \bxi_{i}.$$
Using the distribution of $(\bkappa_{i-1}|\bY_{(i-1)})$ derived in KF3b, and that $\bkappa_{i-1}$ is independent to $\boeta_i$,  we get that 
$$ (\bx_{i} | \bY_{(i-1)}) \sim N(\bmu_1, \bSigma_1).$$


\section{Supplementary figures and tables}
\begin{figure}[h]
\centering
\includegraphics[scale=1]{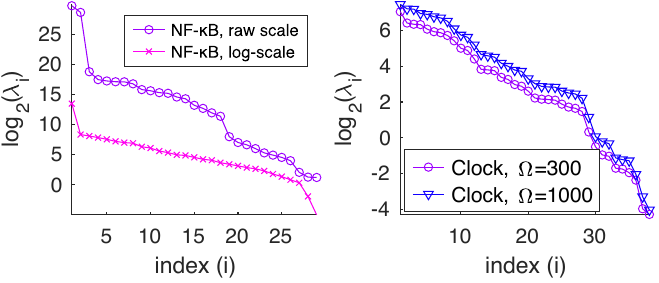}
\caption{\label{appfig:eigRaw} Singular values of the FIM for the pcLNA model of the \NFKB system with parameters in raw and log scale (see legend). The log2 of the singular values is presented here. This is the same setting with Figure 4 in the main document but here all singular values are reported.}
\end{figure}

\begin{figure}
\centering
\includegraphics[scale=0.25]{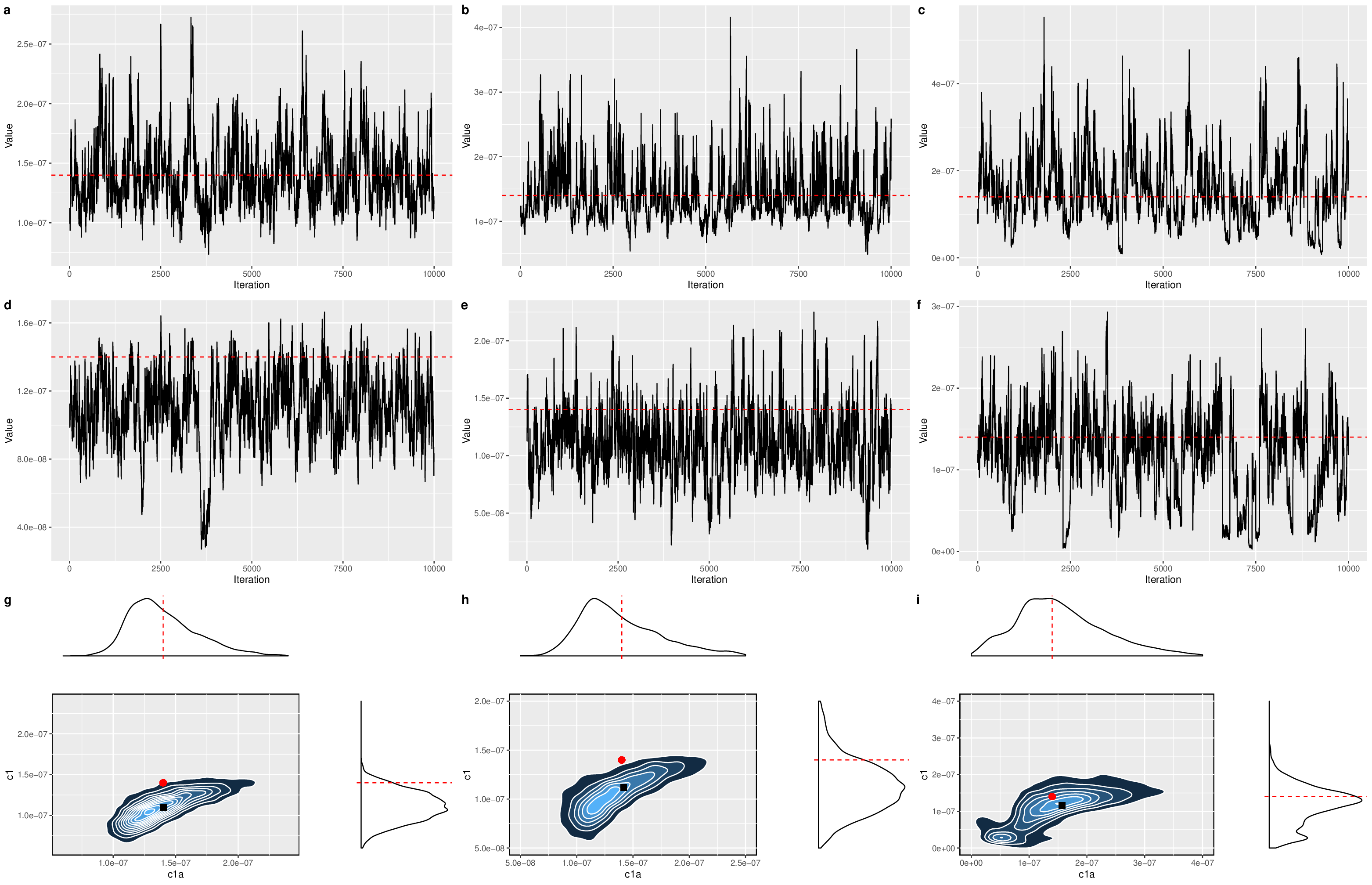}
\caption{\label{suppfig:singular}Posterior inference post-burn-in for parameters $c1a$ and $c1$ in the NF-$\kappa$B system, conducted on the original parameter scale. Top two rows show traceplots for parameter $c1a$ and $c1$ respectively with the bottom row showing bivariate marginal distributions. Each column relates to a single algorithm type (random walk Metropolis, SMMALA and parallel tempered random walk). Red dashed lines/circles correspond to the true parameter values and yellow circles the posterior mean.}
\end{figure}

\begin{figure}
\centering
\includegraphics[width=\textwidth]{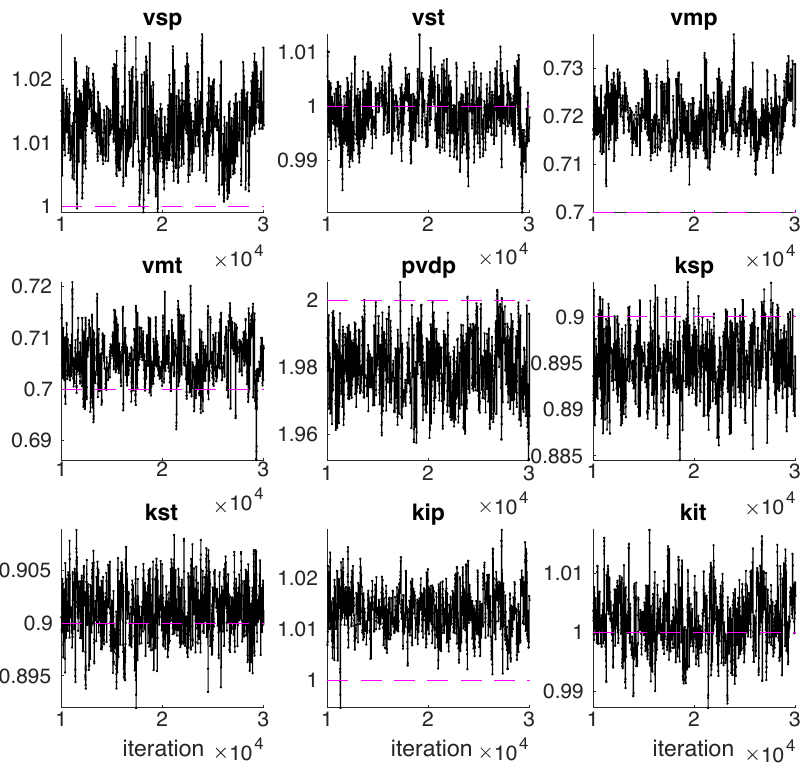}
\caption{\label{fig:dros1000}Traceplots after burnin of $10^4$ for the Drosophila circadian clock system estimating 9 parameters as in Table 3 of the main paper. Dashed horizontal line corresponds to the true parameter value used to simulate the data using the SSA.}
\end{figure}

\begin{figure}
\centering
\includegraphics[width=\textwidth]{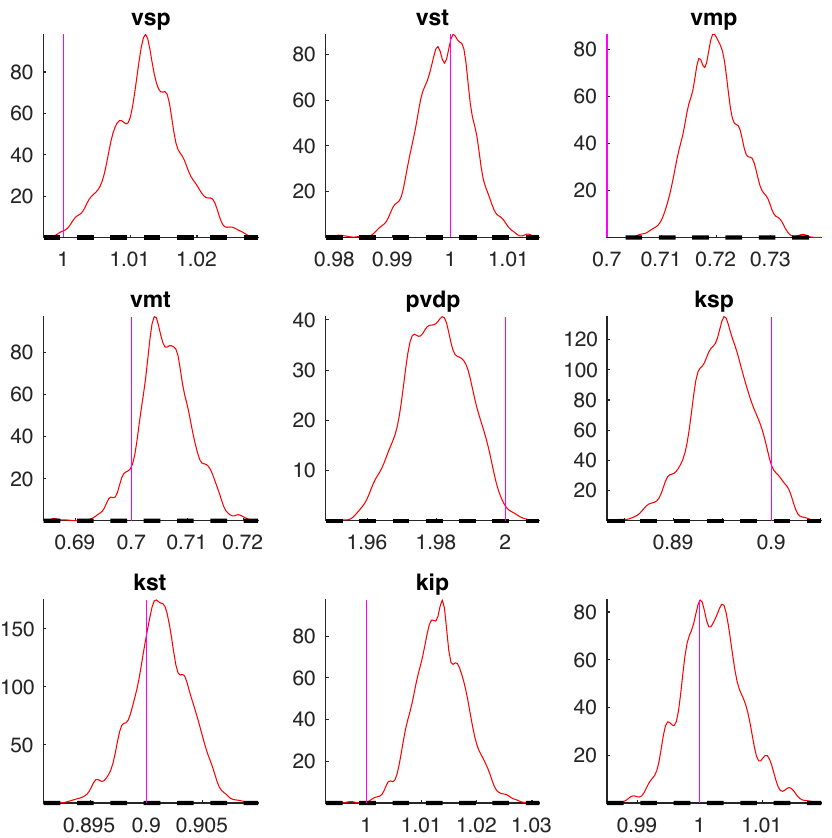}
\caption{\label{fig:drospriorpost}Marginal posterior densities (solid red) with corresponding prior density (dotted black) and true values used for SSA simulations (solid purple) as in Table 3 of the main paper.}
\end{figure}

\begin{figure}
\centering
\includegraphics[width=\textwidth]{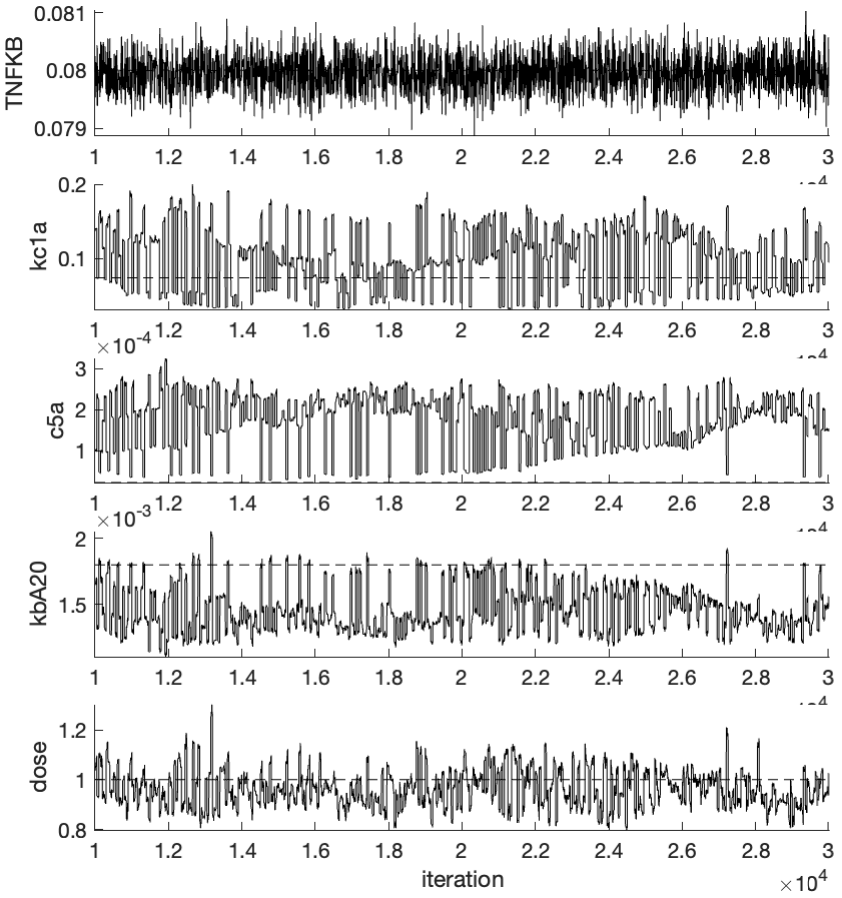}
\caption{\label{nfkbchains_fig10}Traceplots after burnin period of $10^4$ for the PT MCMC run of the \NFKB system estimating 5 parameters. Dashed horizontal line corresponds to the true parameter value used to simulate the data using pcLNA. The Effective Sample Size (ESS) for each chain are, respectively, $1.732$ (TNFKB), $442.2$ (kc1a), $438.7$ (c5a), $441$ (kbA20), $434.1$ (dose).}
\end{figure}

\begin{figure}
\centering
\includegraphics[width=\textwidth]{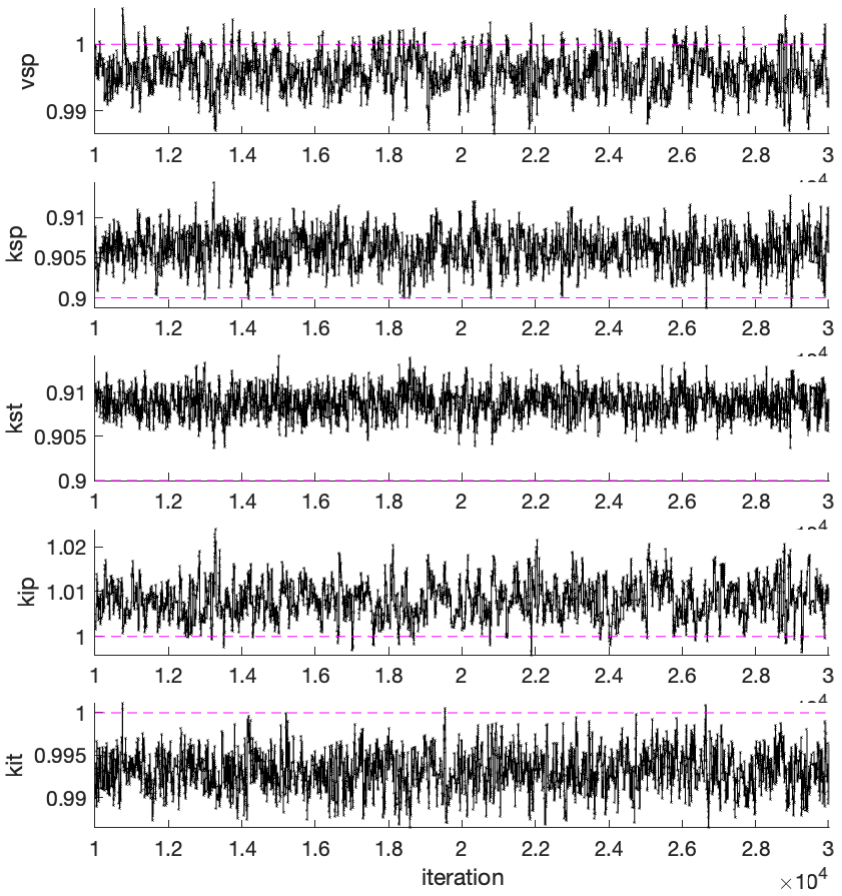}
\caption{\label{high5_chains}Traceplots after burnin period of $10^4$ for the PT MCMC run of the circadian clock system estimating the 5 highest sensitivity parameters. Dashed horizontal line corresponds to the true parameter value used to simulate the data using SSA. The Effective Sample Size (ESS) for each chain are, respectively, $390.1$ (vsp),  $483.1$ (ksp),  $603.3$ (kst),  $404.2$ (kip), $624.7$ (kit).}
\end{figure}

\begin{figure}
\centering
\includegraphics[width=\textwidth]{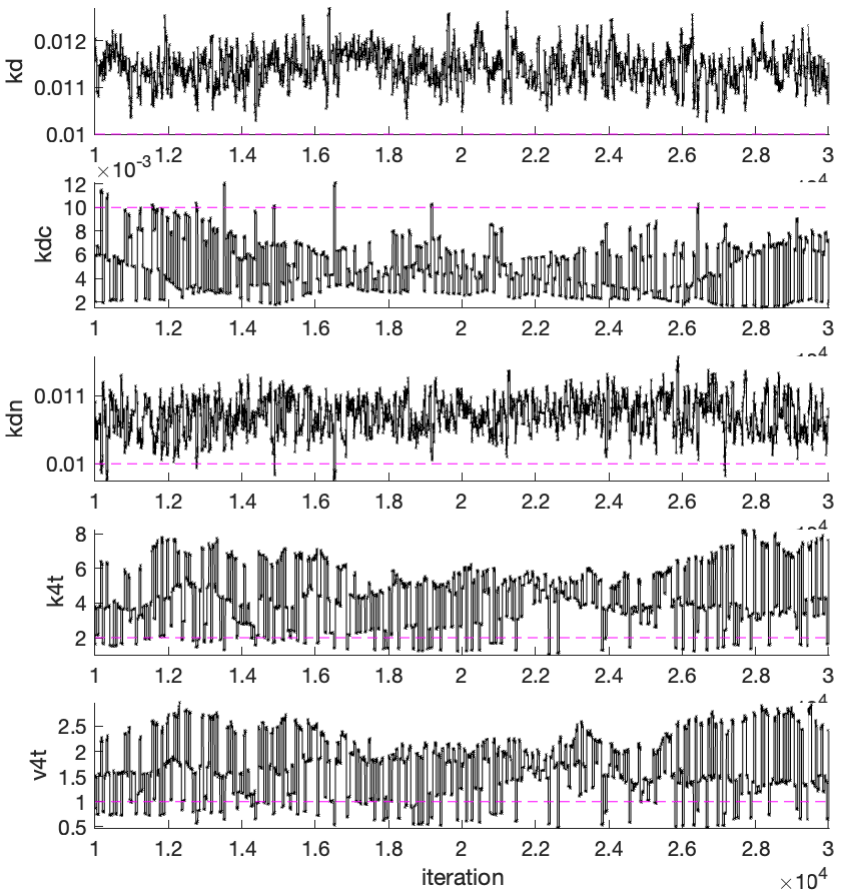}
\caption{\label{low5_chains}Traceplots after burnin period of $10^4$ for the PT MCMC run of the circadian clock system estimating the 5 lowest sensitivity parameters. Dashed horizontal line corresponds to the true parameter value used to simulate the data using SSA. The Effective Sample Size (ESS) for each chain are, respectively, $207.9$ (kd), $456.4$ (kdc), $461.0$ (kdn), $447.6$ (k4t), $452.9$ (v4t).}
\end{figure}

\begin{figure}
\centering
\includegraphics[width=\textwidth]{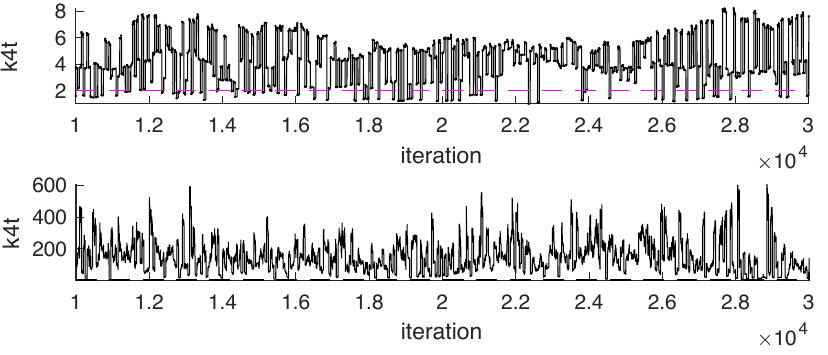}
\caption{\label{k4t_chains}Traceplots for the k4t parameter after burnin period of $10^4$ for the PT MCMC run of the circadian clock system estimating the 5 lowest sensitivity parameters (top) and k4t, v4t parameters (bottom). Dashed horizontal line corresponds to the true parameter value used to simulate the data using SSA. The Effective Sample Size (ESS) for each chain are, respectively, $447.62$ (top) and $429.98$ (bottom).}
\end{figure}

\begin{figure}
\centering
\includegraphics[width=\textwidth]{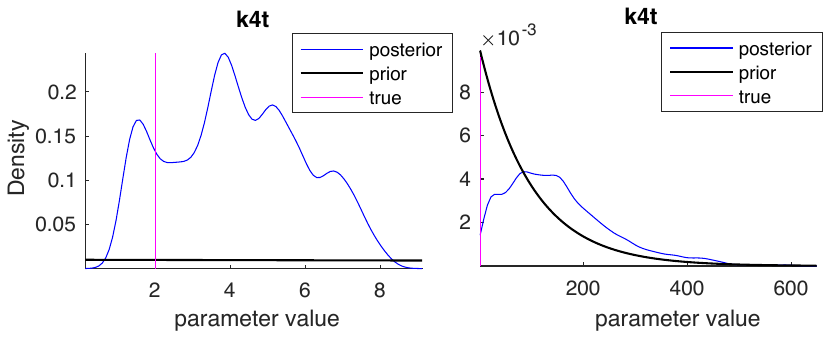}
\caption{\label{k4t_priorVpost}Prior, posterior and the (true) value used for simulation for the parameter k4t of the Drosophila clock system obtained in a PT MCMC run with the 5 lowest sensitivity parameters being estimated (left) and the parameters k4t and v4t being estimated (right).}
\end{figure}

\begin{figure}
\centering
\includegraphics[width=\textwidth]{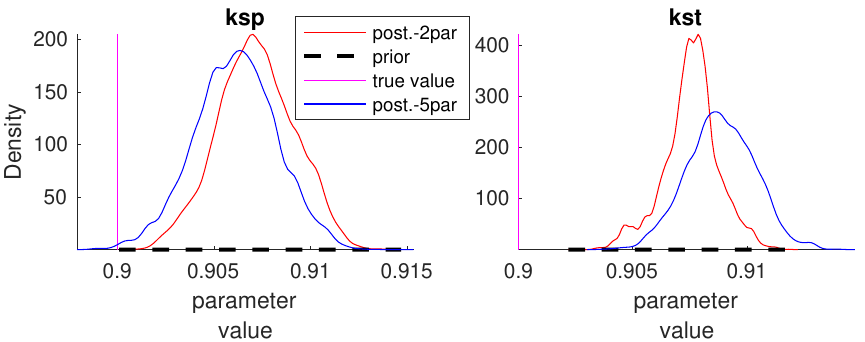}
\caption{\label{high2_priorVpost}Posterior (solid, red line) for 2 parameters estimated (ksp, left, kst, right), prior (dashed, black), the (true) value (solid, purple) used for simulation of the Drosophila clock system, and the posterior when 5 parameters (inluding ksp and kst) are estimated (solid, red) using the PT MCMC. The two posteriors are much more similar to each other than the two posteriors presented in \ref{k4t_priorVpost}.}
\end{figure}
\newpage

\begin{table*}[h!]
\begin{tabular}{cccc}
\hline
Parameter & $\log(\| \bu_k\|)$ & ESS-1 temp. & ESS-6 temp.\\[0.125em]
\hline
TNFKB&9.364&1254&1732.7\\ 
kc1a &2.1696&12.243&442.23\\
c5a  & 1.1835&11.732&438.66\\
kbA20& 4.706&12.969&440.96\\
amp &  5.3117&14.361&434.13\\
sigma& 0&1063.6&1582.3\\
   \hline
\end{tabular}
\caption{\label{sitab:essTab1}Sensitivity coefficients and Effective Sample Size (ESS) for two PT MCMC runs of estimating five parameters of the \NFKB system. The first run used only one temperature value $\beta_c=1$ (equivalent to Random Walk Metropolis MCMC) and the second run 6 temperatures $\beta_c=1,0.9,0.7,0.5,0.3,0.1$.}
\end{table*}

\begin{table*}[h!]
\begin{tabular}{cccc}
\hline
Parameter & $\log(\| \bu_k\|)$ & ESS-4 temp. & ESS-6 temp.\\[0.125em]
\hline
 vsp&4.6252&206.19&   390.09   \\
 ksp&4.8041& 479.9&   483.09   \\
 kst&4.7923&385.71&   603.35   \\
 kip&4.5953&178.93&   404.21   \\
 kit&4.5589&383.51&   624.71   \\
 sigma&0&1342.9&   1526.6   \\
   \hline
\end{tabular}
\caption{\label{sitab:essHigh5}Sensitivity coefficients and Effective Sample Size (ESS) for two PT MCMC runs of estimating five parameters of the circadian clock system. The first run used four temperature values $\beta_c=1,0.5,0.3,0.1$ and the second run 6 temperatures $\beta_c=1,0.9,0.7,0.5,0.3,0.1$.}
\end{table*}

\bibliographystyle{bafiles/ba}
\bibliography{supplementary}